\documentclass[twocolumn]{aastex62}
\usepackage{amsmath}
\usepackage{graphicx}
\usepackage{natbib}
\usepackage[scaled]{helvet}
\usepackage{epsfig}
\usepackage{url}
\bibpunct{(}{)}{;}{a}{}{,}
\newcommand{\kms}{\,km\,s$^{-1}$}
\usepackage{textcomp}
\usepackage{mathpazo}
\usepackage{xspace}
\usepackage{hyperref}
\usepackage{apjfonts}
\usepackage{mathrsfs}
\usepackage{verbatim}

\usepackage{color}
\usepackage[normalem]{ulem}

\newcommand{\bjdtdb}{\ensuremath{\rm {BJD_{TDB}}}}
\newcommand{\feh}{\ensuremath{\left[{\rm Fe}/{\rm H}\right]}}

\newcommand{\teff}{\ensuremath{T_{\rm eff}}\xspace}
\newcommand{\logg}{\ensuremath{\log g}}

\newcommand{\msun}{\ensuremath{\,M_\Sun}}
\newcommand{\rsun}{\ensuremath{\,R_\Sun}}
\newcommand{\lsun}{\ensuremath{\,L_\Sun}}
\newcommand{\mj}{\ensuremath{\,M_{\rm J}}}
\newcommand{\rj}{\ensuremath{\,R_{\rm J}}}

\newcommand{\fave}{\langle F \rangle}
\newcommand{\fluxcgs}{10$^9$ erg s$^{-1}$ cm$^{-2}$}

\newcommand{\Kepler}{{\it Kepler}}
\newcommand{\Ktwo}{{\it K2}}
\newcommand{\loggstar}{\ensuremath{\log{g_\star}}}
\newcommand{\vsini}{\ensuremath{v\sin{I_*}}}
\newcommand{\ms}{\,m\,s$^{-1}$}
\newcommand{\thisstar}{TOI-172\xspace}

\newcommand{\rstar}{\ensuremath{R_{*}}}

\newcommand{\be}{\begin{equation}}
\newcommand{\ee}{\end{equation}}

\begin{document}

\title{An Eccentric Massive Jupiter Orbiting a Sub-Giant on a 9.5 Day Period Discovered in \\ the Transiting Exoplanet Survey Satellite Full Frame Images}

\newcommand{\cfa}{Center for Astrophysics \textbar \ Harvard \& Smithsonian, 60 Garden St, Cambridge, MA 02138, USA}
\newcommand{\umich}{Astronomy Department, University of Michigan, 1085 S University Avenue, Ann Arbor, MI 48109, USA}
\newcommand{\utaustin}{Department of Astronomy, The University of Texas at Austin, Austin, TX 78712, USA}
\newcommand{\MIT}{Department of Physics and Kavli Institute for Astrophysics and Space Research, Massachusetts Institute of Technology, Cambridge, MA 02139, USA}
\newcommand{\MITEPS}{Department of Earth, Atmospheric and Planetary Sciences, Massachusetts Institute of Technology,  Cambridge,  MA 02139, USA}
\newcommand{\uflorida}{Department of Astronomy, University of Florida, 211 Bryant Space Science Center, Gainesville, FL, 32611, USA}
\newcommand{\riverside}{Department of Earth Sciences, University of California,
Riverside, CA 92521, USA}
\newcommand{\usq}{University of Southern Queensland, West St, Darling Heights
QLD 4350, Australia}
\newcommand{\ames}{NASA Ames Research Center, Moffett Field, CA, 94035, USA}
\newcommand{\geneva}{Observatoire de l’Universit\'e de Gen\`eve, 51 chemin des Maillettes,
1290 Versoix, Switzerland}
\newcommand{\uw}{Astronomy Department, University of Washington, Seattle, WA 98195 USA}
\newcommand{\warwick}{Deptartment of Physics, University of Warwick, Gibbet Hill Road, Coventry CV4 7AL, UK}
\newcommand{\warwickceh}{Centre for Exoplanets and Habitability, University of Warwick, Gibbet Hill Road, Coventry CV4 7AL, UK}
\newcommand{\princeton}{Department of Astrophysical Sciences, Princeton University, 4 Ivy Lane, Princeton, NJ, 08544, USA}
\newcommand{\liege}{Space Sciences, Technologies and Astrophysics Research (STAR) Institute, Universit\'e de Li\`ege, 19C All\'ee du 6 Ao\^ut, 4000 Li\`ege, Belgium}
\newcommand{\vanderbilt}{Department of Physics and Astronomy, Vanderbilt University, Nashville, TN 37235, USA}
\newcommand{\fisk}{Department of Physics, Fisk University, 1000 17th Avenue North, Nashville, TN 37208, USA}
\newcommand{\columbia}{Department of Astronomy, Columbia University, 550 West 120th Street, New York, NY 10027, USA}
\newcommand{\toronto}{Dunlap Institute for Astronomy and Astrophysics, University of Toronto, Ontario M5S 3H4, Canada}
\newcommand{\unc}{Department of Physics and Astronomy, University of North Carolina at Chapel Hill, Chapel Hill, NC 27599, USA}
\newcommand{\iac}{Instituto de Astrof\'isica de Canarias (IAC), E-38205 La Laguna, Tenerife, Spain}
\newcommand{\lalaguna}{Departamento de Astrof\'isica, Universidad de La Laguna (ULL), E-38206 La Laguna, Tenerife, Spain}
\newcommand{\louisville}{Department of Physics and Astronomy, University of Louisville, Louisville, KY 40292, USA}
\newcommand{\aavso}{American Association of Variable Star Observers, 49 Bay State Road, Cambridge, MA 02138, USA}
\newcommand{\utokyo}{The University of Tokyo, 7-3-1 Hongo, Bunky\={o}, Tokyo 113-8654, Japan}
\newcommand{\naoj}{National Astronomical Observatory of Japan, 2-21-1 Osawa, Mitaka, Tokyo 181-8588, Japan}
\newcommand{\jstpresto}{JST, PRESTO, 7-3-1 Hongo, Bunkyo-ku, Tokyo 113-0033, Japan}
\newcommand{\astrobiojapan}{Astrobiology Center, 2-21-1 Osawa, Mitaka, Tokyo 181-8588, Japan}
\newcommand{\ctio}{Cerro Tololo Inter-American Observatory, Casilla 603, La Serena, Chile}
\newcommand{\nexsci}{Caltech/IPAC -- NASA Exoplanet Science Institute 1200 E. California Ave, Pasadena, CA 91125, USA}
\newcommand{\ucsc}{Department of Astronomy and Astrophysics, University of
California, Santa Cruz, CA 95064, USA}
\newcommand{\gsfc}{Exoplanets and Stellar Astrophysics Laboratory, Code 667, NASA Goddard Space Flight Center, Greenbelt, MD 20771, USA}
\newcommand{\sgtinc}{SGT, Inc./NASA AMES Research Center, Mailstop 269-3, Bldg T35C, P.O. Box 1, Moffett Field, CA 94035, USA}
\newcommand{\chile}{Center of Astro-Engineering UC, Pontificia Universidad Cat\'olica de Chile, Av. Vicu\~{n}a Mackenna 4860, 7820436 Macul, Santiago, Chile}
\newcommand{\Pontificia}{Instituto de Astrof\'isica, Pontificia Universidad Cat\'olica de Chile, Av.\ Vicu\~na Mackenna 4860, Macul, Santiago, Chile}
\newcommand{\Millennium}{Millennium Institute for Astrophysics, Chile}
\newcommand{\maxplank}{Max-Planck-Institut f\"ur Astronomie, K\"onigstuhl 17, Heidelberg 69117, Germany}
\newcommand{\utdallas}{Department of Physics, The University of Texas at Dallas, 800 West
Campbell Road, Richardson, TX 75080-3021 USA}
\newcommand{\MauryLewin}{Maury Lewin Astronomical Observatory, Glendora, CA 91741, USA}
\newcommand{\umbc}{University of Maryland, Baltimore County, 1000 Hilltop Circle, Baltimore, MD 21250, USA}
\newcommand{\osu}{Department of Astronomy, The Ohio State University, 140 West 18th Avenue, Columbus, OH 43210, USA}
\newcommand{\MITAA}{Department of Aeronautics and Astronautics, MIT, 77 Massachusetts Avenue, Cambridge, MA 02139, USA}
\newcommand{\openu}{School of Physical Sciences, The Open University, Milton Keynes MK7 6AA, UK}
\newcommand{\swarthmore}{Department of Physics and Astronomy, Swarthmore College, Swarthmore, PA 19081, USA}
\newcommand{\seti}{SETI Institute, Mountain View, CA 94043, USA}
\newcommand{\lehigh}{Department of Physics, Lehigh University, 16 Memorial Drive East, Bethlehem, PA 18015, USA}

\newcommand{\torres}{\altaffiliation{Juan Carlos Torres Fellow}}
\newcommand{\sagan}{\altaffiliation{NASA Sagan Fellow}}
\newcommand{\bernoulli}{\altaffiliation{Bernoulli fellow}}
\newcommand{\gruber}{\altaffiliation{Gruber fellow}}
\newcommand{\kavli}{\altaffiliation{Kavli Fellow}}
\newcommand{\peg}{\altaffiliation{51 Pegasi b Fellow}}
\newcommand{\pappalardo}{\altaffiliation{Pappalardo Fellow}}
\newcommand{\hubble}{\altaffiliation{NASA Hubble Fellow}}

\correspondingauthor{Joseph E. Rodriguez} 
\email{joseph.rodriguez@cfa.harvard.edu}

\author[0000-0001-8812-0565]{Joseph E. Rodriguez} 
\affiliation{\cfa}

\author[0000-0002-8964-8377]{Samuel N. Quinn} 
\affiliation{\cfa}

\author[0000-0003-0918-7484]{Chelsea X. Huang} 
\torres
\affiliation{\MIT}

\author[0000-0001-7246-5438]{Andrew Vanderburg} 
\sagan
\affiliation{\utaustin}

\author[0000-0003-4464-1371]{Kaloyan Penev} 
\affiliation{\utdallas}

\author[0000-0002-9158-7315]{Rafael Brahm} 
\affiliation{\chile}
\affiliation{\Pontificia}
\affiliation{\Millennium}

\author[0000-0002-5389-3944]{Andr\'es Jord\'an} 
\affiliation{\Pontificia}
\affiliation{\Millennium}

\author{Mma Ikwut-Ukwa} 
\affiliation{\cfa} 

\author{Shelly Tsirulik} 
\affiliation{\cfa}

\author[0000-0001-9911-7388]{David W. Latham} 
\affiliation{\cfa}

\author[0000-0002-3481-9052]{Keivan G. Stassun} 
\affiliation{\vanderbilt}
\affiliation{\fisk}

\author[0000-0002-1836-3120]{Avi Shporer} 
\affiliation{\MIT}

\author[0000-0002-0619-7639]{Carl Ziegler} 
\affiliation{\toronto}

\author{Elisabeth Matthews} 
\affiliation{\MIT}

\author[0000-0003-3773-5142]{Jason D. Eastman} 
\affiliation{\cfa}

\author[0000-0003-0395-9869]{B. Scott Gaudi} 
\affiliation{\osu}

\author[0000-0001-6588-9574]{Karen A. Collins} 
\affiliation{\cfa}

\author[0000-0002-5169-9427]{Natalia Guerrero} 
\affiliation{\MIT}

\author{Howard M. Relles} 
\affiliation{\cfa}


\author[0000-0001-7139-2724]{Thomas Barclay} 
\affiliation{\gsfc}
\affiliation{\umbc}

\author[0000-0002-7030-9519]{Natalie~M.~Batalha} 
\affiliation{\ucsc}

\author{Perry Berlind} 
\affiliation{\cfa}

\author[0000-0001-6637-5401]{Allyson Bieryla} 
\affiliation{\cfa}

\author[0000-0002-0514-5538]{L. G.~Bouma} 
\affiliation{\princeton}

\author{Patricia T. Boyd} 
\affiliation{\gsfc}


\author[0000-0002-0040-6815]{Jennifer~Burt}
\torres
\affiliation{\MIT}

\author[0000-0002-2830-5661]{Michael L. Calkins} 
\affiliation{\cfa}

\author[0000-0002-8035-4778]{Jessie Christiansen}
\affiliation{\nexsci}

\author[0000-0002-5741-3047]{David R. Ciardi}  
\affiliation{\nexsci}

\author[0000-0001-8020-7121]{Knicole D.\ Col\'on} 
\affiliation{\gsfc}

\author[0000-0003-2239-0567]{Dennis M.\ Conti} 
\affiliation{\aavso}

\author{Ian J.\ M.\ Crossfield} 
\affiliation{\MIT}

\author[0000-0002-6939-9211]{Tansu~Daylan} 
\kavli
\affiliation{\MIT}

\author[0000-0001-7730-2240]{Jason~Dittmann}
\peg
\affiliation{\MIT}

\author[0000-0003-2313-467X]{Diana Dragomir} 
\hubble
\affiliation{\MIT}

\author{Scott~Dynes} 
\affiliation{\MIT}

\author[0000-0001-9513-1449]{N\'estor Espinoza} 
\bernoulli
\gruber
\affiliation{\maxplank}

\author{Gilbert A. Esquerdo} 
\affiliation{\cfa}

\author{Zahra~Essack} 
\affiliation{\MITEPS}


\author[0000-0001-9828-3229]{Aylin~Garcia~Soto} 
\affiliation{\MITEPS}

\author[0000-0002-5322-2315]{Ana~Glidden} 
\affiliation{\MITEPS}
\affiliation{\MIT}

\author[0000-0002-3164-9086]{Maximilian~N.~G{\"u}nther}  
\torres
\affiliation{\MIT}

\author{Thomas Henning} 
\affiliation{\maxplank}

\author[0000-0002-4715-9460]{Jon M. Jenkins} 
\affiliation{\ames}


\author[0000-0003-0497-2651]{John F.\ Kielkopf} 
\affiliation{\louisville}

\author[0000-0002-8781-2743]{Akshata~Krishnamurthy} 
\affiliation{\MITAA}

\author[0000-0001-9380-6457]{Nicholas M. Law} 
\affiliation{\unc}

\author[0000-0001-8172-0453]{Alan M. Levine} 
\affiliation{\MIT}

\author[0000-0003-0828-6368]{Pablo Lewin} 
\affiliation{\MauryLewin}

\author[0000-0003-3654-1602]{Andrew W. Mann} 
\affiliation{\unc}

\author[0000-0003-1447-6344]{Edward H. Morgan} 
\affiliation{\MIT}

\author{Robert L. Morris}
\affiliation{\seti}
\affiliation{\ames}

\author[0000-0002-0582-1751]{Ryan J. Oelkers} 
\affiliation{\vanderbilt}

\author[0000-0001-8120-7457]{Martin Paegert} 
\affiliation{\cfa}

\author[0000-0002-3827-8417]{Joshua Pepper} 
\affiliation{\lehigh}

\author[0000-0003-1309-2904]{Elisa V. Quintana} 
\affiliation{\gsfc}

\author{George R. Ricker} 
\affiliation{\MIT}

\author[0000-0002-4829-7101]{Pamela~Rowden} 
\affiliation{\openu}

\author[0000-0002-6892-6948]{Sara Seager} 
\affiliation{\MIT}
\affiliation{\MITEPS}
\affiliation{\MITAA}

\author[0000-0001-8128-3126]{Paula Sarkis} 
\affiliation{\maxplank}

\author{Joshua E. Schlieder} 
\affiliation{\gsfc}

\author{Lizhou~Sha}
\affiliation{\MIT}


\author{Andrei Tokovinin} 
\affiliation{\ctio}

\author[0000-0002-5286-0251]{Guillermo Torres} 
\affiliation{\cfa}

\author{Roland K. Vanderspek} 
\affiliation{\MIT}

\author[0000-0001-6213-8804]{Steven~Villanueva~Jr.} 
\pappalardo
\affiliation{\MIT}

\author{Jesus Noel Villase\~{n}or} 
\affiliation{\MIT}

\author[0000-0002-4265-047X]{Joshua N. Winn} 
\affiliation{\princeton}

\author{Bill Wohler} 
\affiliation{\seti}
\affiliation{\ames}

\author{Ian Wong} 
\peg
\affiliation{\MITEPS}

\author{Daniel A. Yahalomi} 
\affiliation{\cfa}

\author[0000-0003-1667-5427]{Liang~Yu} 
\affiliation{\MIT}

\author[0000-0002-4142-1800]{Zhuchang~Zhan} 
\affiliation{\MITEPS}

\author[0000-0002-4891-3517]{George Zhou} 
\hubble
\affiliation{\cfa}

\shorttitle{TOI-172}
\shortauthors{Rodriguez et al.}

\begin{abstract}
We report the discovery of TOI-172 b from the Transiting Exoplanet Survey Satellite ({\it TESS}) mission, a massive hot Jupiter transiting a slightly evolved G-star with a 9.48-day orbital period. This is the first planet to be confirmed from analysis of only the {\it TESS} full frame images, because the host star was not chosen as a two minute cadence target. From a global analysis of the {\it TESS} photometry and follow-up observations carried out by the {\it TESS} Follow-up Observing Program Working Group, TOI-172 (TIC 29857954) is a slightly evolved star with an effective temperature of $T_{\rm eff}$ =$5645\pm50$~K, a mass of $M_{\star}$ = $1.128^{+0.065}_{-0.061}$ $M_{\odot}$, radius of $R_{\star}$ = $1.777^{+0.047}_{-0.044}$ $R_{\odot}$, a surface gravity of $\log$ $g_{\star}$ = $3.993^{+0.027}_{-0.028}$, and an age of $7.4^{+1.6}_{-1.5}$ Gyr. Its planetary companion (TOI-172 b) has a radius of $R_{\rm P}$ = $0.965^{+0.032}_{-0.029}$ $R_{\rm J}$, a mass of $M_{\rm P}$ = $5.42^{+0.22}_{-0.20}$ $M_{\rm J}$, and is on an eccentric orbit ($e = 0.3806^{+0.0093}_{-0.0090}$). TOI-172 b is one of the few known massive giant planets on a highly eccentric short-period orbit. Future study of the atmosphere of this planet and its system architecture offer opportunities to understand the formation and evolution of similar systems.


\end{abstract}


\section{Introduction} 
In only three decades, the field of exoplanets has rapidly expanded from its infancy to one of the largest and fastest research areas in astrophysics. This is largely due to the success of both ground-based and space-based efforts to discover new planets using the transit and radial velocity (RV) techniques. With the confirmation of thousands of
new planets and the identification of a few thousand more candidates, no survey has been more influential to the field than the \Kepler\ mission \citep{Borucki:2010}. As the \Kepler\ and re-purposed \Ktwo\ \citep{Howell:2014} missions have completed, we are now entering the next major chapter in the field of exoplanets with the recent launch of the Transiting Exoplanet Survey Satellite ({\it TESS}, \citealp{Ricker:2015}).

\begin{figure*}[!ht]
\vspace{0.3in}
\centering\includegraphics[width=0.95\linewidth, trim = 0 5.8in 0 0]{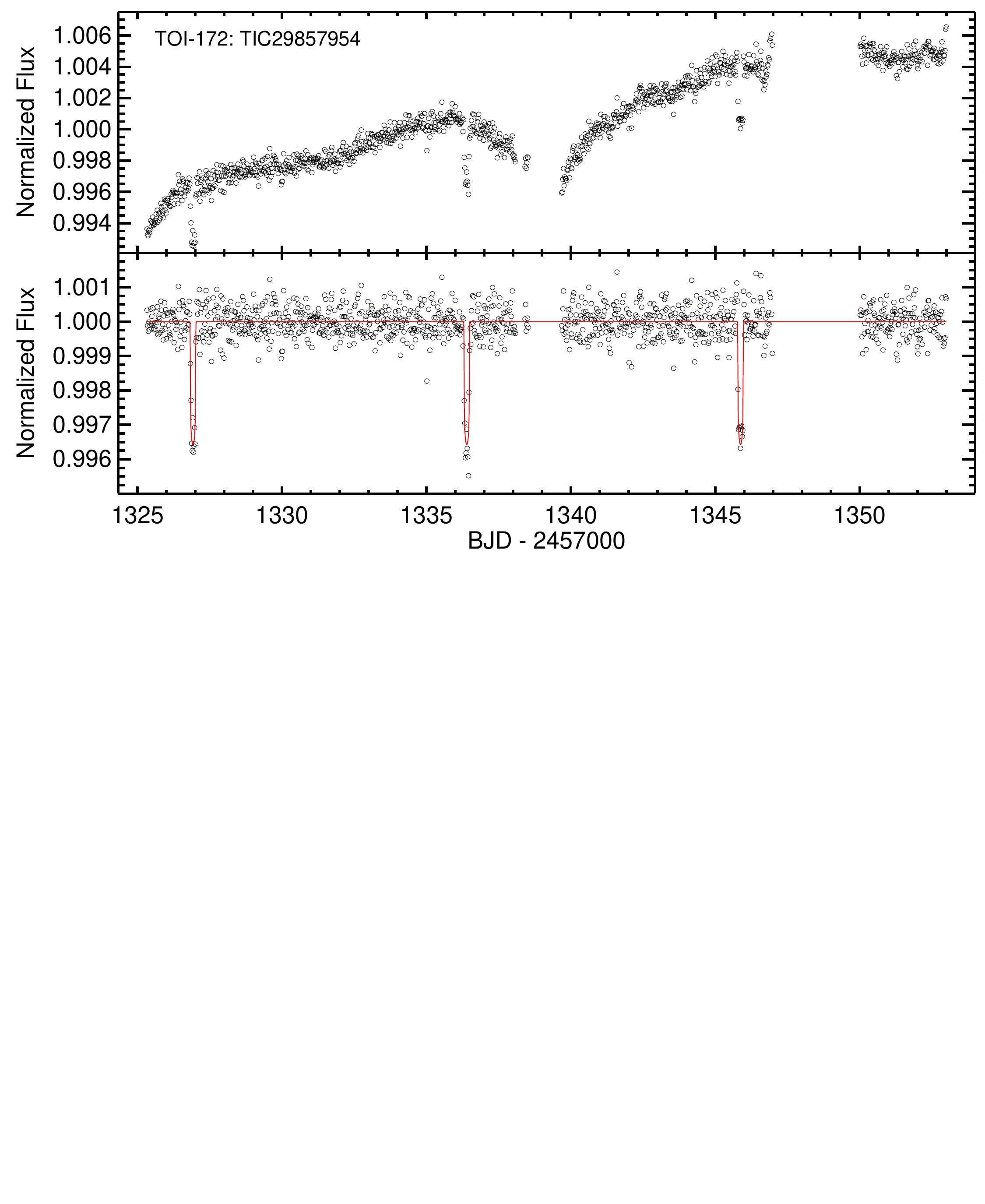}
\caption{(Top) {\it TESS} 30-minute cadence light curve of \thisstar. (Bottom) The flattened final {\it TESS} light curve used in the EXOFASTv2 fit. The observations are plotted in open black circles, and the best fit model from EXOFASTv2 is plotted in red. The gap in the middle is due to the gap between TESS orbits. The data between BJD$_{\rm TDB}$- 2457000 of 1347 to 1350 were removed due to high scatter caused when the spacecraft thrusters are fired to reorient the spacecraft and allow the reaction wheels to spin down. There is also a small 8-hour gap on BJD$_{\rm TDB}$- 2457000 = 1338 due to an asteroid crossing the aperture for \thisstar.}
\label{figure:LC}
\end{figure*}

Interestingly, we are still attempting to understand one of the first types of planets ever discovered, hot Jupiters. It is commonly believed that close-in giant planets formed farther out in the protoplanetary disk and, through various mechanisms, migrated inward. These highly irradiated, Jovian-sized planets orbit with periods $\le10$\,days, and typically do not have nearby planetary companions \citep{Steffen:2012, Huang:2016}, suggesting that they might disrupt planet formation and the orbits of any existing inner planets as they move inward. However, the discovery of two small planets bracketing the known hot Jupiter, WASP-47b \citep{Becker:2015}, suggests that some giant planets can migrate in a dynamically quiet manner or even form in-situ \citep{Huang:2016,Batygin:2016}. It has been found that giant planets discovered in more distant orbits tend to have companions \citep[e.g.][]{Knutson:2014, Huang:2016}.  This possibly supports the idea that their longer orbit allows them to form alongside smaller planets in different parts of the inner disk. 


If planetary migration occurs through the gas disk, it must take place during the first $\sim$10 Myr while the gas is still around, and is expected to result in low-eccentricity orbits \citep{Haisch:2001, DAngelo:2003}. However, migration may commonly be influenced by gravitational interactions with other planets or stars. These interactions can increase the planet's orbital eccentricity (known as ``High Eccentricity Migration" (HEM)) and lead to tidal interactions at close approach to the host star that shrink and circularize the orbit \citep{Rasio:1996, Wu:2003, Fabrycky:2007,Nagasawa:2011, Wu:2011}. 
For a Jupiter analogue orbiting a Sun-like star on a period of 0.5 to 10 days, the circularization timescale can range from a few million years to over a hundred billion years depending on semi-major axis (see equation 2 in \citealp{Adams:2006}). Therefore, only long period hot Jupiters (5--10 days) would retain any primordial eccentricity if HEM is the underlying mechanism because they would not have had enough time to circularize. This class of ``dynamically young" giant planets, for which the circularization timescales are longer than the system's current age \citep[also referred to as "tropical Jupiters";][]{Yu:2018}, offers an opportunity to gain insight into the mechanisms governing hot Jupiter evolution. Previous studies have tried to place constraints on hot Jupiter migration mechanisms by analyzing the eccentricities and orbital architectures of these systems. For example, the orbits of dynamically young hot Jupiters tend to be more eccentric on average, as would be expected if at least a fraction of them have undergone eccentric migration \citep{Quinn:2014, Bonomo:2017}. At the same time, the paucity of highly eccentric migrating Jupiters places an upper limit on the prevalence of HEM in the production of these systems \citep{Dawson:2015}. The presence of additional giant planets exterior to hot Jupiters but inside the ice line is hard to reconcile with migration via HEM \citep{Schlaufman:2016}, though trends with host star metallicity hint that disk migration and {\it subsequent} planet-planet scattering could account for much of the hot Jupiter population \citep{Dawson:2013}. Indeed, it appears that no single migration channel can produce the known population; a recent review of the relevant literature suggests that the combination of two such mechanisms might be able to explain the observations \citep{Dawson:2018}. Additional study of the dynamically young planets---and their orbital architectures---can refine our understanding of how these migration mechanisms work together to produce the population of giant planets that we observe.


In this paper, we present the discovery in TESS full frame images leading to follow-up photometry, and precision radial velocity measurements of a dynamically young, massive Jupiter in a $\sim$9.5-day eccentric orbit (0.38) around a sub-giant. Additionally, the evolutionary state of \thisstar\ provides a reliable age. The paper is organized in the following way. We present all available observations of \thisstar in \S\ref{Obs} (Table \ref{tbl:LitProps} presents the available information on \thisstar from the literature). Our global analysis of all available observations using EXOFASTv2 is described in \S\ref{sec:GlobalModel}. We discuss \thisstar\ b in the context of all known planets in \S\ref{sec:discussion}, presenting prospects on future follow-up. We summarize our conclusions in \S\ref{sec:conclusion}.


\begin{table}
\scriptsize
\setlength{\tabcolsep}{2pt}
\centering
\caption{Literature and Measured Properties for \thisstar}
\begin{tabular}{llcc}
  \hline
  \hline
Other identifiers\dotfill & \\
\multicolumn{3}{c}{TIC 29857954} \\
\multicolumn{3}{c}{TYC 6932-00301-1} \\
\multicolumn{3}{c}{2MASS J21063165-2641333}\\
\hline
\hline
Parameter & Description & Value & Source\\
\hline 
$\alpha_{J2000}$\dotfill	&Right Ascension (RA)\dotfill & 21:06:31.65& 1	\\
$\delta_{J2000}$\dotfill	&Declination (Dec)\dotfill & -26:41:34.29& 1	\\
\\
B$_T$\dotfill			&Tycho B$_T$ mag.\dotfill & 12.211 $\pm$ 0.203		& 2	\\
V$_T$\dotfill			&Tycho V$_T$ mag.\dotfill & 11.382 $\pm$ 0.125		& 2	\\
${\rm G}$\dotfill     & Gaia $G$ mag.\dotfill     &11.193$\pm$0.02& 3,4\\
${\rm T}$\dotfill     & TESS mag.\dotfill     &10.711$\pm$0.019& 4\\
\\
J\dotfill			& 2MASS J mag.\dotfill & 10.135  $\pm$ 0.03	& 5, 6	\\
H\dotfill			& 2MASS H mag.\dotfill & 9.825 $\pm$ 0.03	    &  5, 6	\\
K$_S$\dotfill			& 2MASS ${\rm K_S}$ mag.\dotfill & 9.722 $\pm$ 0.02&  5, 6	\\
\\
\textit{WISE1}\dotfill		& \textit{WISE1} mag.\dotfill & 9.673 $\pm$ 0.03 & 7	\\
\textit{WISE2}\dotfill		& \textit{WISE2} mag.\dotfill & 9.718 $\pm$ 0.03 &  7 \\
\textit{WISE3}\dotfill		& \textit{WISE3} mag.\dotfill &  9.763 $\pm$ 0.052& 7	\\
\textit{WISE4}\dotfill		& \textit{WISE4} mag.\dotfill & 8.529 $\pm$ 0.516 &  7	\\
\\
$\mu_{\alpha}$\dotfill		& Gaia DR2 proper motion\dotfill & -4.711 $\pm$ 0.094 & 3,4 \\
                    & \hspace{3pt} in RA (mas yr$^{-1}$)	&  \\
$\mu_{\delta}$\dotfill		& Gaia DR2 proper motion\dotfill 	&  -54.25 $\pm$ 0.069 &  3,4 \\
                    & \hspace{3pt} in DEC (mas yr$^{-1}$) &  \\
$v\sin{i_\star}$\dotfill &  Rotational velocity (\kms) \hspace{9pt}\dotfill &  5.1 $\pm$ 0.5 & \S\ref{sec:TRES}\\
$\feh$\dotfill &   Metallicity \hspace{9pt}\dotfill & 0.14 $\pm$ 0.08 & \S\ref{sec:TRES} \\
$\teff$\dotfill &  Effective Temperature (K) \hspace{9pt}\dotfill &  5640 $\pm$ 50  &  \S\ref{sec:TRES}  \\
$\log{g_{\star}}$\dotfill &  Surface Gravity (cgs)\hspace{9pt}\dotfill &  3.97 $\pm$ 0.1  &  \S\ref{sec:TRES} \\
$\pi$\dotfill & Gaia Parallax (mas) \dotfill & 2.972 $\pm$ 0.06$^{\dagger}$ &  3,4 \\
$RV$\dotfill & Systemic radial \hspace{9pt}\dotfill  & $-6.247\pm0.081$  & \S\ref{sec:TRES} \\
     & \hspace{3pt} velocity (\kms)  & \\
$d$\dotfill & Distance (pc)\dotfill & $336.47\pm 6.79^{\dagger}$ & 3,4 \\
$U^{*}$\dotfill & Space Velocity (\kms)\dotfill & $26.24 \pm 0.46$  & \S\ref{sec:uvw} \\
$V$\dotfill       & Space Velocity (\kms)\dotfill & $-71.52 \pm 1.68$ & \S\ref{sec:uvw} \\
$W$\dotfill       & Space Velocity (\kms)\dotfill & $ -1.31 \pm 0.27$ & \S\ref{sec:uvw} \\
\hline
\end{tabular}
\begin{flushleft}
 \footnotesize{ \textbf{\textsc{NOTES:}}
 $\dagger$ Values have been corrected for the -0.82 $\mu$as offset as reported by \citet{Stassun:2018}.\\
 $*$ $U$ is in the direction of the Galactic center. \\
 References are: $^1$\citet{Cutri:2003},$^2$\citet{Hog:2000},$^2$\citet{Gaia:2016},$^3$\citet{Gaia:2018},$^4$\citet{Stassun:2018_TIC}, $^5$\citet{Cutri:2003}, $^6$\citet{Skrutskie:2006}, $^7$\citet{Zacharias:2017}
}
\end{flushleft}
\label{tbl:LitProps}
\end{table}

\section{Observations and Archival Data}
\label{Obs}

\subsection{{\it TESS} Photometry}
\label{sec:TESS}
\thisstar\ fell on CCD~4 of Camera~1 of the {\it TESS} spacecraft during its first sector of observations (2018 July 25 -- August 22), but it was not pre-selected for two-minute cadence observations. After the data were downloaded from the spacecraft, we processed the calibrated 30-minute cadence full frame images \citep{Jenkins:2016} with the MIT Quick Look Pipeline (QLP, C. Huang et al., in preparation). The QLP is a lightweight tool for rapidly producing light curves and identifying transits in \textit{all} stars observed by {\it TESS}, not just those selected for two-minute cadence observations. The QLP extracts photometry by summing the flux within moving circular apertures (following \citealt{Huang:2015}), after using the nebulosity filter\footnote{developed by Irwin (2010):{\url{http://www.ukirt.hawaii.edu/publications/newsletter/ukirtnewsletter2010spring.pdf}}} to remove scattered background light from the images). After producing light curves, the QLP searches for transits by calculating a Box-Least-Squares periodogram (BLS, \citealp{kovacs:2002}), implementing high-pass filtering and BLS period spacing following \citet{Vanderburg:2016b}.  We detected a single repeating transit signal around \thisstar\ with a period of 9.48 days, a duration of 4.71 hours, and a flat-bottomed shape (see Figure \ref{figure:LC}). We notified the community of the discovery via the MIT {\it TESS} Alerts portal\footnote{\url{https://tess.mit.edu/alerts/}} \citep{ricker_tess_alerts_2018}. 

\begin{figure*}[ht!]
	\vspace{.0in}
	\includegraphics[width=6.7in]{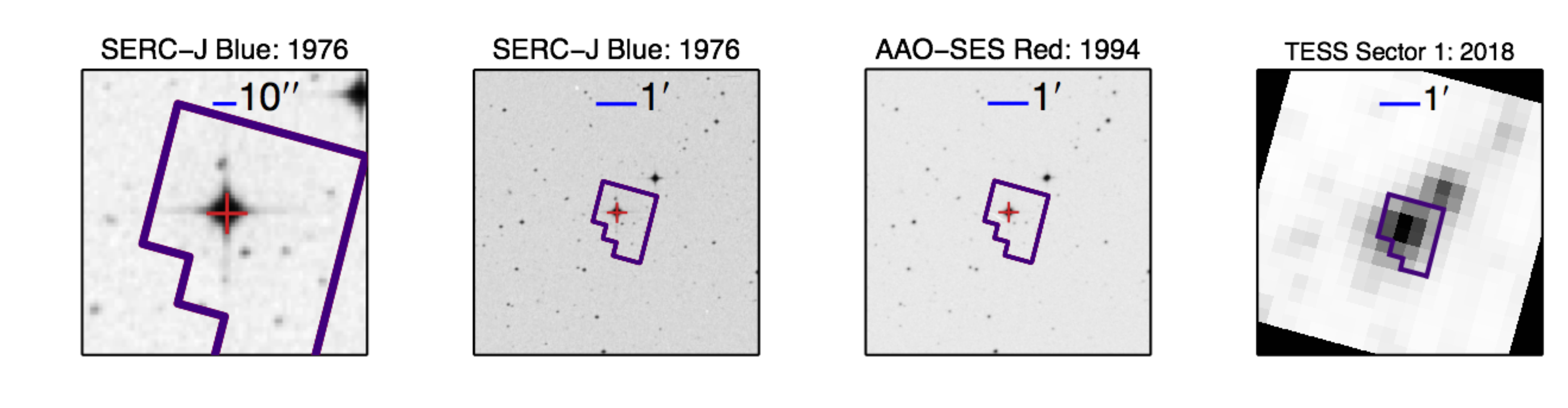}
	\caption{Archival imaging of \thisstar from the ESO/SERC Southern Sky Atlas (SERC-J; taken in 1976, first and second panels) and the Anglo-Australian Observatory Second Epoch Survey (AAO-SES; 1994, 3rd panel). (4th panel) The {\it TESS} image of \thisstar from Sector 1. The outline on each image corresponds to the final chosen aperture used to extract the {\it TESS} light curve and the blue horizontal bar shows the image scale.}
	\label{fig:apertures} 
\end{figure*}

Upon the public release of the processed Sector 1 full frame images, we attempted to improve the light curve by extracting photometry from a variety of differently shaped stationary photometric apertures. After some experimentation, and using archival images from the ESO/SERC Southern Sky Atlas (SERC-J; taken in 1975) and the Anglo-Australian Observatory Second Epoch Survey (AAO-SES; 1993) to check for any additional stars nearby, we settled upon the irregularly shaped aperture shown in Figure \ref{fig:apertures}. The light curve extracted from this aperture balanced high photometric precision with minimal contamination from a nearby 12th magnitude star (TIC 29857959) and minimal systematics related to {\it TESS}'s ``momentum dumps'', when the spacecraft thrusters are fired to reorient the spacecraft and allow the reaction wheels to be spun down. We compared the transit depths from the light curve extracted with the QLP and our simple aperture photometry method, and found consistent results. We proceeded in our analysis using the light curve produced with simple aperture photometry, as it had slightly better photometric precision. We manually removed 8 hours of data (BJD$_{\rm TDB}$- 2457000 = 1338.4125 to 1338.0792) contaminated by an asteroid passing through the photometric aperture, and we clipped 4$\sigma$ outliers from the light curve (see Figure \ref{figure:LC}). The corresponding {\it TESS} light curve was flattened by using a spline fit with breakpoints every 0.5 days to divide out the best-fit stellar variability \citep{Vanderburg:2014}.

\begin{figure}[!ht]
\vspace{0.3in}
\centering\includegraphics[width=0.99\linewidth, trim = 0.0 2.2in 0.0 0in]{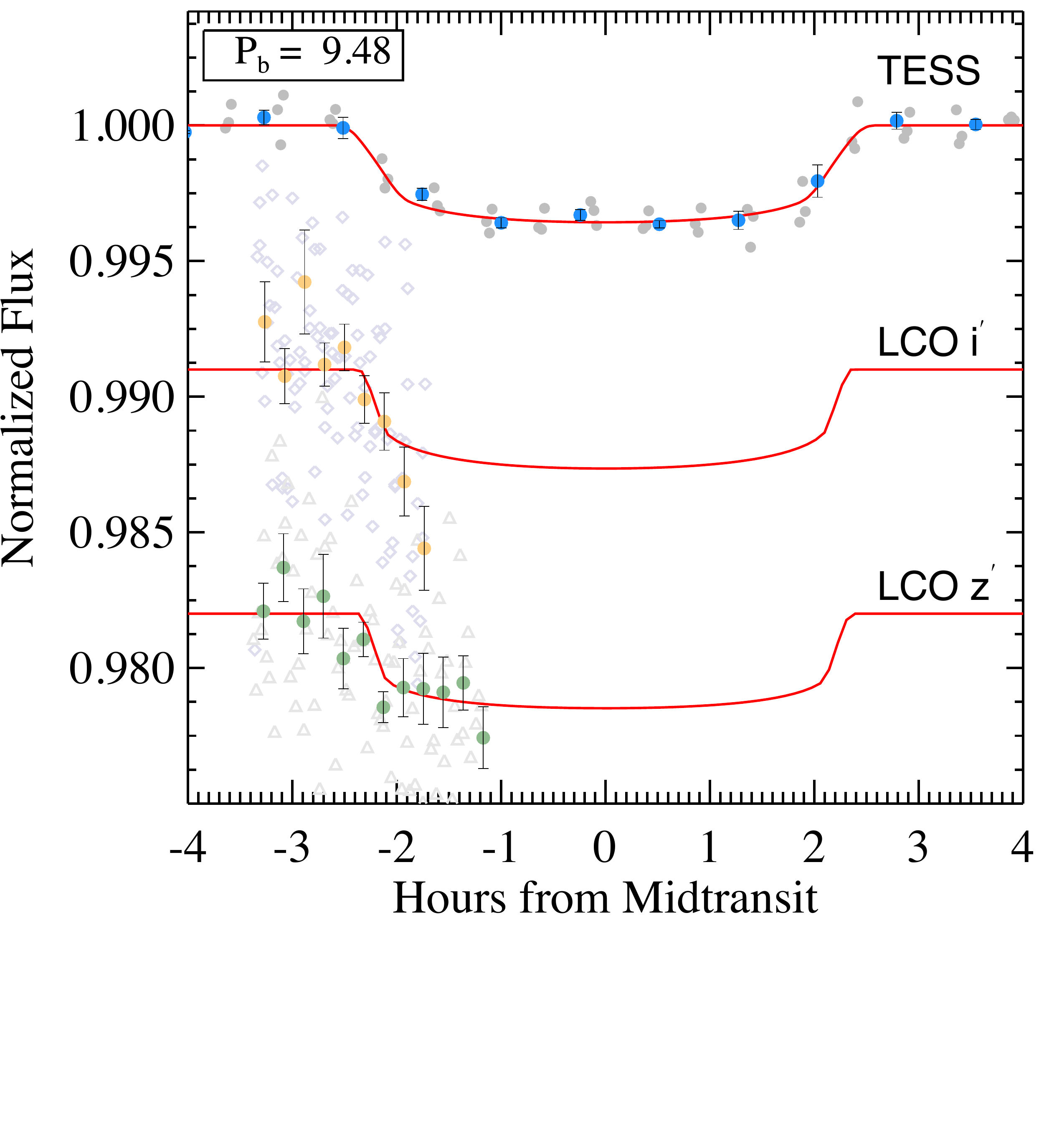}
\caption{ The phase-folded corrected {\it TESS} (blue), LCO i$^\prime$ (yellow), and LCO z$^\prime$ (green) light curves for \thisstar b. The full light curves are shown with filled circles ({\it TESS}), open diamonds (LCO i$^\prime$) and open triangles (LCO z$^\prime$) and the binned points are shown in color with error bars. The bin sizes are 45 minutes for {\it TESS} and 11 minutes for the LCO observations. The red line corresponds to the final EXOFASTv2 transit model. }
\label{figure:LCplanets}
\end{figure}

\subsection{Ground-based Photometry from the {\it TESS} Follow-up Observing Program Working Group}
\label{sec:sg1}
To rule out false positives, better constrain the ephemeris of \thisstar\ b, and measure the depth of the transit, we obtained two photometric transit follow-up observations using the Las Cumbres Observatory (LCO) telescope network \citep{Brown:2013}\footnote{https://lco.global/}. To predict the next possible transit events for \thisstar that were observable, we used the \texttt{TAPIR} software package \citep{Jensen:2013}. We used the \texttt{AstroImageJ} astronomical observation analysis software to reduce all follow-up photometric observations and perform aperture photometry to extract the light curves. On UT 2018 September 22, we observed the transit of \thisstar\ b in the SDSS i$^{\prime}$ filter using the 0.4m LCO telescope located at the Cerro Tololo Inter-American Observatory (CTIO) in Chile. The 0.4m telescopes are equipped with SBIG STX6303 cameras that have a $19\arcmin \times29\arcmin$ field-of-view, and a 0.57$\arcsec$ pixel scale. On UT 2018 October 11, we observed the transit of \thisstar\ b in the z$^{\prime}$ filter on the 1.0m telescope at the McDonald Observatory in Fort Davis, Texas. The 1.0m telescope has a Sinistro camera with a $16.5\arcmin\times26.5\arcmin$ FOV and a pixel scale of 0.389$\arcsec$ pixel$^{-1}$. In each case, an ingress of the transit of \thisstar\ b was observed on the target star. In both observations, only an ingress was observable and the exposure time was 50s. These observations are consistent with what was observed by {\it TESS} (see Figure \ref{figure:LCplanets}). Therefore, the fading events are localized to within 15$\arcsec$ of \thisstar.

\begin{figure}
	\centering\vspace{.0in}
	\includegraphics[width=1\linewidth, page=2, trim={2.5cm 13cm 8.5cm 8cm}, clip]{TIC29857954_rv.pdf}
	\includegraphics[width=1\linewidth, page=1, trim={2.5cm 13cm 8.5cm 8cm}, clip]{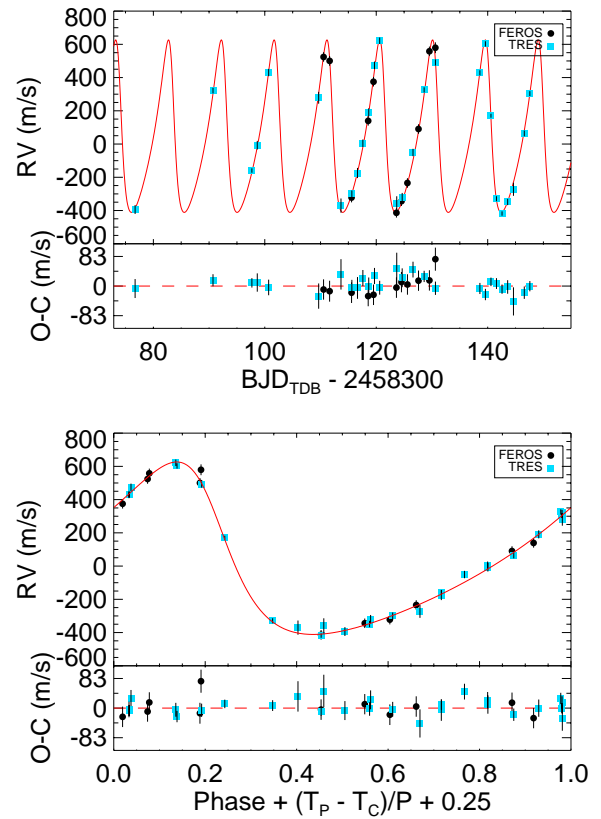}
	\caption{(Top) Radial velocity measurements from FEROS (black) and TRES (blue). (Bottom) The radial velocity measurements are phase-folded to the best determined period by EXOFASTv2, 9.477 days. The EXOFASTv2 model is shown in red and the residuals to the best fit are shown below each plot.}
	\label{fig:TRES_RVs} 
\end{figure}

\subsection{TRES Spectroscopy} \label{sec:TRES}
Spectra of \thisstar were obtained on 27 occasions with a resolving power of R$\sim$44000 using the Tillinghast Reflector Echelle Spectrograph \citep[TRES;][]{furesz:2008}\footnote{\url{http://www.sao.arizona.edu/html/FLWO/60/TRES/GABORthesis.pdf}} mounted on the 1.5m Tillinghast Reflector at the Fred L. Whipple Observatory (FLWO) on Mt. Hopkins, AZ. For a description of the reduction and radial velocity (RV) extraction pipeline, see \citet{Buchhave:2010}. Our procedure differed only in the generation of the template used for cross-correlation. We derived relative RVs by cross-correlating against the strongest spectrum, and we shifted and median-combined the spectra to produce a high-SNR template spectrum. We then cross-correlated each observed spectrum against that template to produce our final relative RVs, which are given in Table \ref{tab:rv} and shown in Figure \ref{fig:TRES_RVs}. Bisector spans were calculated for the TRES RVs using the technique described in \citet{Torres:2007}. There are no correlations between the bisector spans and the measured RV values and no scatter in the bisectors beyond their uncertainties (which are small compared to the RV variation), supporting the premise that \thisstar\ is being periodically transited or eclipsed. We also derive the absolute RVs via cross-correlation against synthetic templates created using Kurucz model atmospheres \citep{Kurucz:1992}. We calculate the instrumental zero-point through nightly monitoring of RV standards, which we place on the absolute RV scale of \citet{Nidever:2002}. Using these observations, we determine the absolute center-of-mass velocity of \thisstar to be -6.247$\pm$0.081 \kms (consistent with the absolute RV from Gaia DR2 of -5.89 $\pm$ 0.67). 

To determine the stellar parameters of \thisstar, we analyzed the TRES spectra using the Stellar Parameter Classification (SPC) analysis package \citep{Buchhave:2012}. From this analysis, we estimated the effective temperature, metallicity, surface gravity, and rotational velocity of \thisstar to be: $\teff$ = 5640 $\pm$ 50 K, $\loggstar$ = 3.97 $\pm$ 0.1, [$m/H$] = 0.14 $\pm$ 0.08, and $\vsini$ = 5.1 $\pm$ 0.5 km s$^{-1}$. 
We use the \teff\ and \feh\ as a prior in the EXOFASTv2 global fit (see \S\ref{sec:GlobalModel}).

\begin{deluxetable}{l l l l l}[bt]
\tabletypesize{\scriptsize}
\tablecaption{Relative Radial Velocities for \thisstar \label{tab:rv}}
\tablewidth{0pt}
\tablehead{
\colhead{\bjdtdb} & \colhead{RV (m s$^{-1}$)} & \colhead{$\sigma_{RV}$ (m s$^{-1}$)} & \colhead{Bisectors} & \colhead{Instrument}
}
\startdata
2458410.551705 &  -5716.4 & 7.4 &   -1.0$\pm$10.0  & FEROS \\
2458411.636884 &  -5739.9 & 9.9 &    6.0$\pm$13.0  & FEROS \\
2458415.573206 &  -6562.4 & 8.6 &   -4.0$\pm$11.0  & FEROS \\
2458418.552974 &  -6100.9 & 7.1 &   -18.0$\pm$10.0  & FEROS \\
2458419.511487 &  -5865.5 & 7.9 &    38.0$\pm$11.0  & FEROS \\
2458423.625864 &  -6654.1 & 8.3 &   -15.0$\pm$11.0  & FEROS \\
2458424.531225 &  -6584.4 & 7.0 &    8.0$\pm$10.0  & FEROS \\
2458425.597896 &  -6474.7 & 7.8 &   -7.0$\pm$11.0  & FEROS \\
2458427.583262 &  -6149.8 & 7.4 &   -8.0$\pm$10.0  & FEROS \\
2458429.542763 &  -5682.3 & 7.4 &    13.0$\pm$10.0  & FEROS \\
2458430.614729 &  -5660.9 & 17.1&    35.0$\pm$19.0  & FEROS \\
\hline
2458376.730770 & -593.6 & 27.0 & 13.3$\pm$19.0  & TRES \\
2458390.710107 & 122.0 & 17.6 &  7.5$\pm$14.7 & TRES \\
2458397.687254 & -359.0 & 14.0 & 23.7$\pm$21.2  & TRES \\
2458398.649461 & -206.5 & 26.2 & 16.5$\pm$18.4  & TRES \\
2458400.688662 & 230.9 & 22.4 & 8.5$\pm$21.9  & TRES \\
2458409.670769 & 78.7 & 35.8 & -56.7$\pm$33.4  & TRES \\
2458413.668975 & -569.8 & 43.0 & -123.3$\pm$51.7  & TRES \\
2458415.631951 & -499.7 & 21.2 & -5.3$\pm$22.6  & TRES \\
2458416.639897 & -373.8 & 30.9 & 13.6$\pm$29.0  & TRES \\
2458417.601186 & -194.6 & 23.1 & 1.0$\pm$21.6  & TRES \\
2458418.659549 & -8.6 & 24.7 & 14.1$\pm$28.5  & TRES \\
2458419.685623 & 274.1 & 23.1 & -10.4$\pm$35.5  & TRES \\
2458420.614450 & 422.1 & 20.0 & 16.4$\pm$16.1  & TRES \\
2458423.681839 & -559.6 & 46.0 & -9.8$\pm$23.3  & TRES \\
2458424.653259 & -520.3 & 26.9 & 25.1$\pm$25.3  & TRES \\
2458426.597813 & -250.4 & 22.6 & 10.5$\pm$25.6  & TRES \\
2458428.594704 & 126.1 & 16.5 & 29.0$\pm$15.2  & TRES \\
2458430.627687 & 291.2 & 19.2 & 37.342$\pm$2.3  & TRES \\
2458438.602799 & 228.7 & 16.3 & 25.6$\pm$11.2  & TRES \\
2458439.594972 & 403.1 & 16.6 & -3.7$\pm$11.5  & TRES \\
2458440.586306 & -29.0 & 12.8 & 1.5$\pm$13.9  & TRES \\
2458441.577056 & -528.6 & 15.8 & 30.8$\pm$13.1  & TRES \\
2458442.582383 & -617.8 & 13.8 & 21.4$\pm$14.7  & TRES \\
2458443.586964 & -548.2 & 19.5 & -15.5$\pm$22.6  & TRES \\
2458444.630963 & -471.7 & 39.9 & -67.3$\pm$45.2  & TRES \\
2458446.575996 & -135.6 & 18.0 & -33.4$\pm$18.1  & TRES \\
2458447.572304 & 104.0 & 15.1 & 29.4$\pm$15.9  & TRES \\
\enddata
\end{deluxetable}

\subsection{FEROS Spectroscopy}

We also obtained 9 R=48000 spectra of \thisstar between UT 2018 October 19 and November 5 using the FEROS spectrograph \citep{Kaufer:99} mounted on the 2.2m MPG telescope at La Silla observatory in Chile. Each spectrum achieved a signal-to-noise ratio of $\sim 60$--$100$ per spectral resolution element with exposure times of 600 sec. The instrumental drift was determined via comparison with a simultaneous fiber illuminated with a ThAr+Ne lamp. The data were processed with the \texttt{CERES} suite of echelle pipelines \citep{Brahm:2017}, which produce radial velocities and bisector spans in addition to reduced spectra.

\subsection{High Resolution Imaging}
\label{sec:AO}
\begin{figure*}[!ht]
\centering 
\includegraphics[height=2.5in]{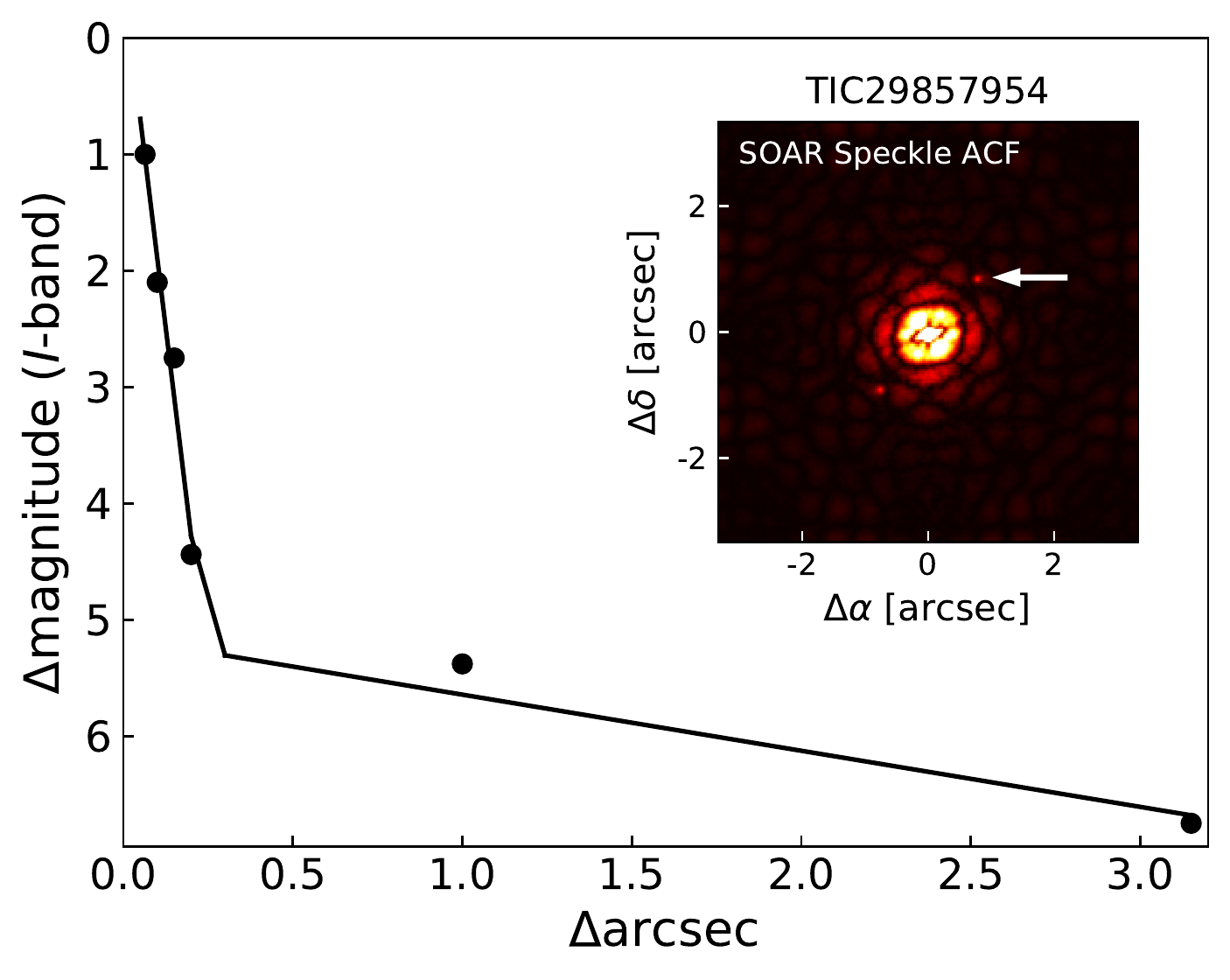}
\includegraphics[trim = 0 0 0 0.15in,height=2.5in]{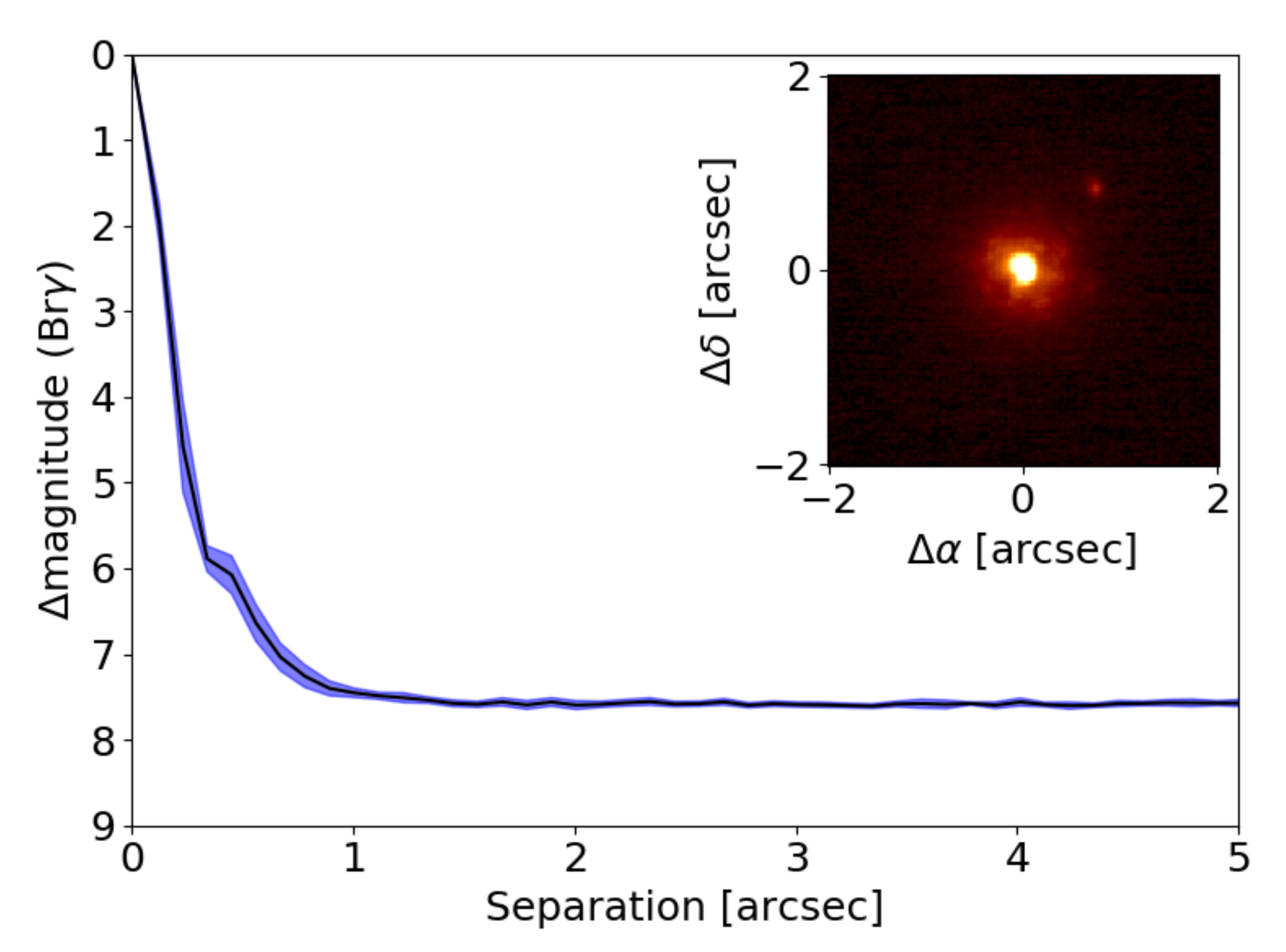}
\caption{{\it Left}: The $I$-band auto-correlation function from Speckle using SOAR. The 5-$\sigma$ contrast curve for \thisstar\ is shown by the black points. The black solid line is the linear fit to the data for separations $<$ 0.2$\arcsec$ and $>$0.2$\arcsec$. The auto-correlation function is shown within the contrast curve plot. {\it Right}: The Br$\gamma$-band AO image and 5-$\sigma$ contrast curve for \thisstar. The faint companion is detected in both data sets; in the speckle ACF, the white arrow points to the position of the visual companion, which is mirrored in the ACF by the speckle processing. }
\label{fig:SOAR}
\end{figure*}

The relatively large 21\arcsec\ pixels of {\it TESS} can result in photometric contamination from nearby sources. These must be accounted for to help rule out astrophysical false positives, such as background eclipsing binaries, and to correct the estimated planetary radius, initially derived from the diluted transit in a blended light curve \citep{Ciardi:2015, Ziegler:2018}. We searched for close companions to \thisstar\ with speckle imaging on the 4.1-m Southern Astrophysical Research (SOAR) telescope \citep{Tokovinin:2018} on UT 2018 September 25, and again in better conditions on UT 2018 October 21. 
We also obtained adaptive optics (AO) images of the target on UT 2018 November 14 using Gemini/NIRI. For these observations, 9 science frames with exposure time 11 seconds each were collected, with the telescope dithered between each frame. For a subset of the frames, the raw data showed signs of stripping, and so we discarded these frames and combined only the 6 good frames for the analysis. We flat field and sky subtract the frames, using a sky background constructed by median combining the dithered images, and then align and combine the images. Figure \ref{fig:SOAR} shows the $5\sigma$ detection limits along with the AO image and speckle auto-correlation function.

A nearby star was detected in both the speckle and the AO observations. The object is measured at a separation of $1.104\arcsec$ and an I-band contrast of 4.9 mag in the speckle images, and at a separation of $1.099\arcsec$ and a Br$\gamma$ contrast of 4.5 mag in the AO images. This would result in a 371 au projected separation if the companion is at the same distance as \thisstar. 
To test this assumption, we use the broadband photometry in these two bands for \thisstar (2MASS $K_s$ = 9.722 and USNO $I^{\prime}$ = 10.50, \citealp{Monet:2003}) and the measured contrasts for the companion ($\Delta$Br$\gamma$ = 4.5 and $\Delta$I = 4.9) to derive the $I-K_s$ colors to be 0.78 and 1.18, respectively. Using these colors and the MIST stellar evolution models \citep{Dotter:2016, Choi:2016, Paxton:2011, Paxton:2013, Paxton:2015} at a log(age) = 9.9 and a solar metallicity (the nearest isochrone grid for \thisstar), we estimate a photometric distance to \thisstar to be 318 pc. This is in close agreement with the {\it Gaia} distance (336.47 pc). However, assuming the companion is on the main sequence, we estimate its distance to be 1092 pc. When using other MIST isochrone grids near the one adopted here, we only see a small change in the derived photometric distances, not nearly enough to explain the large difference measured between \thisstar and the visual companion. This discrepancy suggests that the visual companion is likely a background object, and not gravitationally bound to the planet host. More data are required to confirm this conclusion, either in the form of more photometry to further characterize the SED of the visual companion, or additional astrometric measurements that confirm whether the two stars share common proper motion.


The nearby star is not in Gaia DR2 or the {\it TESS} input catalog, and consequently was not accounted for in the contamination correction for \thisstar. It would take a 28.5\% deep eclipse of the nearby faint companion to cause the blended depth seen in our aperture for \thisstar. The high contrast between the two stars significantly reduces the possibility that the nearby star is a background eclipsing binary resulting in a false positive planetary transit signal, as does subsequent radial velocity follow-up. We know that such a faint companion is unable to significantly affect the RVs of \thisstar because its contribution to the line profile is so small. While it is true that even a faint companion could affect the RVs slightly (even if it is below the noise level of the CCF), this would only be at the level of m/s, not hundreds of m/s (TOI-172b k = 517 \ms, see the analysis of blended CCFs in \citealp{Buchhave:2011}). Therefore, the spectroscopy proves that the planetary companion orbits our target.  Assuming the primary star is the planet host, the additional flux from the nearby star results in only a negligible correction upwards to the initially derived planet radius ($\sim$0.5\%). We account for the blending from this nearby companion in our global fit (see \S\ref{sec:GlobalModel}).

\begin{figure}[!ht]
\centering 
\includegraphics[trim = 1.3in 0.6in 1in 1.15in,width=\columnwidth]{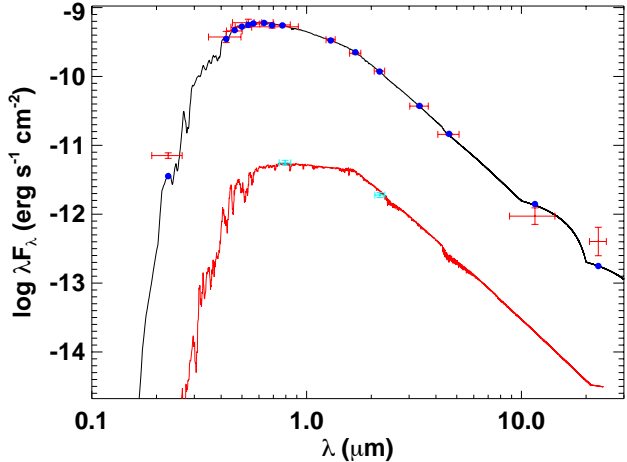}
\caption{The two-component SED fit for \thisstar. The blue points are the predicted integrated fluxes and the red points are the observed values at the corresponding passbands. The cyan points correspond to the I-band flux of the nearby companion observed by SOAR ($\Delta$I-band) and the Br$\gamma$ flux observed by Gemini. The width of the bandpasses are the horizontal red error bars and the vertical errors represent the 1$\sigma$ uncertainties. The final model fit is shown by the solid line for \thisstar\ (black) and its companion (red). }
\label{fig:sed_fit}
\end{figure}

\subsection{SED Analysis}
\label{sec:SED}
Due to the presence of a nearby visual companion (see Figure \ref{fig:SOAR}), we are unable to simultaneously fit the Spectral Energy Distribution (SED) within the EXOFASTv2 global analysis. Instead, we fit the combined SED of the two stars separately from the join transit and RV analysis. The companion is blended in each of the broadband photometric observations. From our analysis of the speckle high-resolution imaging, we know that the nearby companion has an I-band contrast of 4.9 mag and a Br$\gamma$ contrast of 4.5 mag. Using the available photometric observations (see Table \ref{tbl:LitProps}), we fit the broadband SED of \thisstar\ spanning 0.2--20~$\mu$m (Figure~\ref{fig:sed_fit}). Assuming both stars have the same $A_V$, we use the $\Delta$I and $\Delta$Br$\gamma$ contrasts to fit an SED to the nearby companion. Each flux measurement is fit using the stellar atmosphere models of \citet{Kurucz:1992}. The distance for \thisstar\ is adopted from the measured \textit{Gaia} parallax and we use the SPC determined \teff, \loggstar, and \feh\ as Gaussian priors on the fit. The only free parameter is the extinction ($A_V$) which is constrained at its upper bound by the maximum permitted line-of-sight extinction from \citet{Schlegel:1998}. Our final best-fit SED for \thisstar\ has a reduced $\chi^2$ of 1.7 and an extinction $A_V = 0.08 \pm 0.04$, and is shown in Figure \ref{fig:sed_fit}. We integrated the best-fit SED to determine the unextincted bolometric flux (correcting for the contamination of the companion) received at Earth, $F_{\rm bol} = 8.16\pm 0.19 \times 10^{-10}$ erg~s$^{-1}$~cm$^{-2}$. Using the \textit{Gaia} parallax (corrected for the systematic offset reported by \citealp{Stassun:2018}) combined with the adopted \teff\ from this analysis, we are able to measure the radius of \thisstar to be $R_\star = 1.787  \pm 0.049$~\rsun, after accounting for the presence of the nearby companion seen in our high-resolution imaging. We use this determined $R_\star$ as a prior for the EXOFASTv2 global analysis (see \S\ref{sec:GlobalModel}). Using our two-component SED fit, we determine the flux contribution of the nearby companion to be 0.91\% (TESS), 0.46\% (g$^{\prime}$), and 1.07\% (z$^{\prime}$). We note that the contribution from the companion would correspond to a change in the measured {\it TESS} transit depth $<$1$\sigma$.

\subsection{Location in the Galaxy, UVW Space Motion, and Galactic Population}
\label{sec:uvw}
\thisstar\ is located at $\alpha_{\rm J2000} =21^h06^m31\fs65$ and $\delta_{\rm J2000} =-26\arcdeg41\arcmin34\farcs29$, and from Gaia DR2 the parallax is $2.89\pm0.06$~mas (applying the correction from \citealp{Stassun:2018}), corresponding to a distance of $336.47\pm 6.79$~pc ignoring the Lutz-Kelker bias, which can cause measured parallaxes to be larger than they are due to the assumption that the number of observable stars increases as you go farther out \citep{Lutz:1973}. This results in \thisstar\ being 217.6 pc below the Galactic plane. Combining the \textit{Gaia} DR2 proper motions of $(\mu_{\alpha},\mu_{\delta})=(-4.711 \pm 0.094, -54.25 \pm 0.069)~{\rm mas~yr}^{-1}$, the \textit{Gaia} parallax, and the absolute radial velocity as determined from the TRES spectroscopy of $-6.25 \pm 0.081 ~{\rm km~s^{-1}}$, we determine the three-dimensional Galactic space motion of $(U,V,W)=(26.24 \pm 0.46, -71.52 \pm 1.68, -1.31 \pm 0.27)~{\rm km~s^{-1}}$, where positive $U$ is in the direction of the Galactic center. We adopt the \citet{Coskunoglu:2011} determination of the solar motion with respect to the local standard of rest. The large asymmetric drift (large negative V velocity) of the host star, combined with its relatively large vertical height below the plane, suggests that the star could potentially be a member of the thick disk.  Indeed, \thisstar\ has a 43.9\% chance of being in the thin disk according to the classification scheme of \citet{Bensby:2003}. However, this conclusion is somewhat contraindicated by the slightly super-solar metallicity of the host star.  We suggest a measurement of the star's detailed elemental abundances (in particular [$\alpha$/Fe]) could clarify the Galactic population to which this star belongs.


\begin{figure*}
\vspace{.0in}
\centering\includegraphics[width=.5\linewidth]{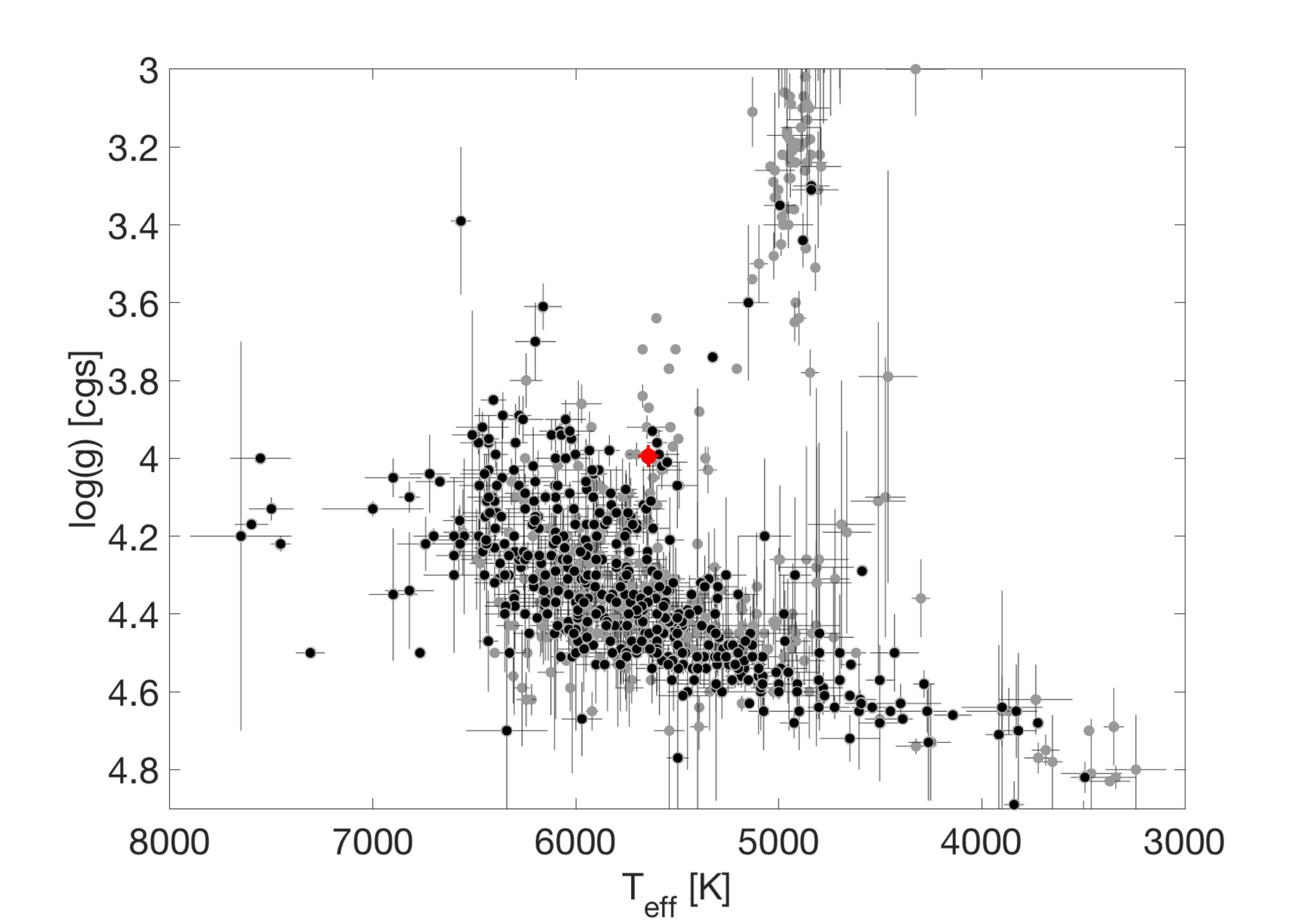}\includegraphics[width=.38\linewidth, angle = 90, trim = 0.5in 0 0 0]{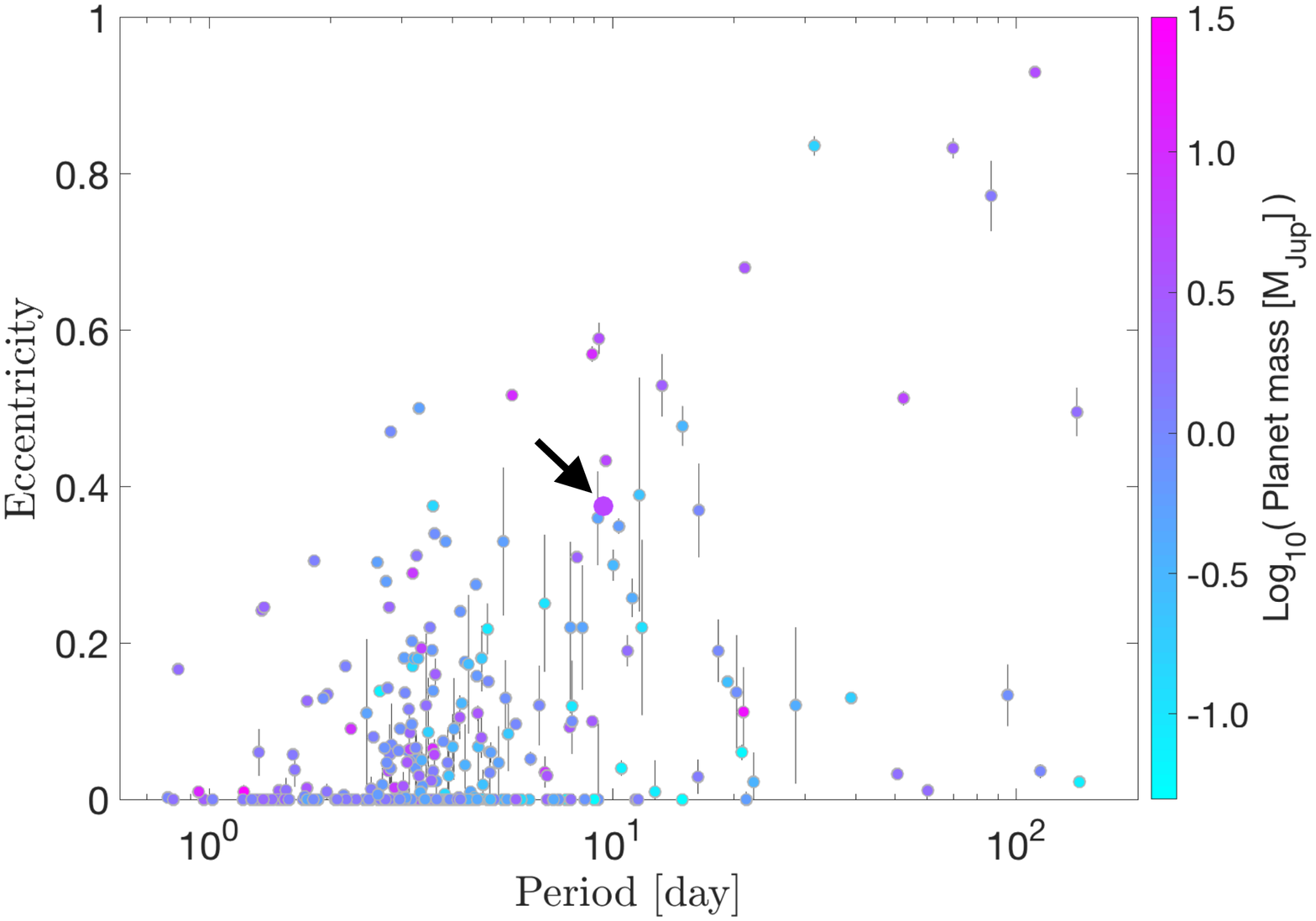}
\caption{(Left) \logg\ and \teff\ of all known stars with transiting (black) and radial velocity (grey) discovered exoplanets. \thisstar\ b is shown in red. (Right) Period and eccentricity of all known exoplanets color coded by log($M_P$).  \thisstar\ b is identified by the larger data point and the black arrow. The data behind these figures were downloaded on UT 2018 December 21 from the NASA Exoplanet Archive \citep{Akeson:2013}.
}
\label{fig:discussion} 
\end{figure*}

\section{EXOFAST\lowercase{v}2 Global Fit for \thisstar} 
\label{sec:GlobalModel}

We use the EXOFASTv2 modeling suite \citep{Eastman:2013,Eastman:2017} to perform a simultaneous fit of the available photometric and spectroscopic observations to gain a full understanding of the \thisstar system. EXOFASTv2 is heavily based on the original EXOFAST modeling suite \citep{Eastman:2013} but provides flexibility in allowing the user to simultaneously fit the SED, RV observations from multiple instruments, and an arbitrary number of planets. We simultaneously fit the full frame image {\it TESS} light curve (see \S\ref{sec:TESS} and Figure \ref{figure:LC}), accounting for the effect of the 30-minute cadence smearing on the light curve, the follow-up ingresses observed by LCO, and the radial velocity observations from TRES and FEROS (see Figure \ref{fig:TRES_RVs}). From our Speckle observations and two-component SED analysis, we found that the nearby companion 1.1$\arcsec$ from \thisstar contributes 0.91\%, 0.46\%, and 1.07\% of the total flux of the system in the {\it TESS}, g$^{\prime}$, and z$^{\prime}$ band-passes. To properly deblend the {\it TESS} and follow-up observations from the previously unknown companion, we include these flux contribution with a 5\% error as Gaussian priors in the EXOFASTv2 global fit. This error has no influence on the determined results. 

Because accurate TESS pixel response function (PRF) models are not yet available, we did not attempt to deblend the TESS light curve from contaminating flux from TIC 29857959, the 12th magnitude star 75 arcseconds northwest of \thisstar\ (see the discussion of this object in Section \ref{sec:TESS} and Figure \ref{fig:apertures}). We did, however, confirm that the neighbor's contaminating flux does not significantly dilute the transit depth of \thisstar\ b using several methods. First, we extracted the light curve of \thisstar\ from even smaller apertures than the one shown in Figure \ref{fig:apertures}, and found that decreasing the aperture size had no effect on the depths of the transits (empirically showing dilution is not an important factor). We also estimated the local TESS PRF by examining TESS images of the nearby isolated bright star (TIC 29857846). Inspection of these images showed that in this region of the TESS field of view, about 75\% of the total flux falls within one pixel of the peak of the PRF, and virtually all of the flux falls within about 6 pixels of the peak. The photometric aperture for \thisstar\ covers about 10\% of the detector area within six pixels of TIC 29857959, so only about 2.5\% of the neighboring star's total flux contaminates \thisstar's aperture. Since TIC 29857959 is about 1.3 magnitudes fainter than \thisstar, the contamination from TIC 29857959 should only be about 1\% the total flux in the aperture, much smaller than the uncertainties on the depth of the transit, and therefore negligible for the transit fitting\footnote{The uncertainty on the transit depth is about 4\%, so 1\% dilution affects the measured depth by much less than 1$\sigma$.}. Finally, we note that this estimate of about 1\% contamination from TIC 29857959 is consistent with the contamination estimated by version 7 of the TESS Input Catalog using pre-launch estimates of the PRF. 

To characterize the host star within the fit, we use the MESA Isochrones and Stellar Tracks (MIST) stellar evolution models \citep{Dotter:2016, Choi:2016, Paxton:2011, Paxton:2013, Paxton:2015}. We enforce Gaussian priors on \teff\ (5640$\pm$50 K) and \feh\ (0.14$\pm$0.08) from the SPC analysis of the TRES spectra (see \S\ref{sec:TRES}). We also place a Gaussian prior on \rstar\ of 1.783$\pm$0.049 \rsun\ from the two-component SED analysis that included the Gaia DR2 parallax (see \S\ref{sec:SED}). The final determined system parameters for \thisstar are shown in Tables \ref{tab:exofast_stellar} \& \ref{tab:exofast}. Our determined R$_{\star}$ is larger than what was listed in the TESS input catalog (TIC) because version 7 of the TIC did not have a Gaia parallax for \thisstar\ and relied on color relations that are unable to distinguish between dwarfs and sub giants. 

\begin{table}
\small
\centering
\caption{Median values and 68\% confidence interval for global model of \thisstar}
\begin{tabular}{llcccc}
  \hline
  \hline
Parameter & Units & Values & & & \\
\hline
\multicolumn{2}{l}{Stellar Parameters:}&\smallskip\\
~~~~$M_*$\dotfill &Mass (\msun)\dotfill &$1.128^{+0.065}_{-0.061}$\\
~~~~$R_*$\dotfill &Radius (\rsun)\dotfill &$1.777^{+0.047}_{-0.044}$\\
~~~~$L_*$\dotfill &Luminosity (\lsun)\dotfill &$2.89^{+0.19}_{-0.18}$\\
~~~~$\rho_*$\dotfill &Density (cgs)\dotfill &$0.286^{+0.022}_{-0.023}$\\
~~~~$\log{g}$\dotfill &Surface gravity (cgs)\dotfill &$3.993^{+0.027}_{-0.028}$\\
~~~~$T_{\rm eff}$\dotfill &Effective Temperature (K)\dotfill &$5645\pm50$\\
~~~~$[{\rm Fe/H}]$\dotfill &Metallicity (dex)\dotfill &$0.148^{+0.079}_{-0.080}$\\
~~~~$[{\rm Fe/H}]_{0}^\dagger$\dotfill &Initial Metallicity \dotfill &$0.172^{+0.074}_{-0.078}$\\
~~~~$Age$\dotfill &Age (Gyr)\dotfill &$7.4^{+1.6}_{-1.5}$\\
~~~~$EEP^\ddagger$\dotfill &Equal Evolutionary Point \dotfill &$456.0^{+3.5}_{-6.8}$\\
\hline
\end{tabular}
\begin{flushleft} 
  \footnotesize{ 
    \textbf{\textsc{NOTES:}}
$^\dagger$The initial metallicity is the metallicity of the star when it was formed.
$^\ddagger$The Equal Evolutionary Point corresponds to static points in a stars evolutionary history when using the MIST isochrones and can be a proxy for age. See \S2 in \citet{Dotter:2016} for a more detailed description of EEP.
               }
 \end{flushleft}
\label{tab:exofast_stellar}
\end{table}
\begin{table*}
\scriptsize
\centering
\caption{Median values and 68\% confidence interval for global model of \thisstar}
\begin{tabular}{llcccc}
  \hline
  \hline
Parameter & Description (Units) & Values & & & \\
\hline
~~~~$P$\dotfill &Period (days)\dotfill &$9.47725^{+0.00064}_{-0.00079}$\\
~~~~$R_{\rm P}$\dotfill &Radius (\rj)\dotfill &$0.965^{+0.032}_{-0.029}$\\
~~~~$T_{\rm C}$\dotfill &Time of conjunction (\bjdtdb)\dotfill &$2458326.9190\pm0.0017$\\
~~~~$T_{\rm 0}^\dagger$\dotfill &Optimal conjunction Time (\bjdtdb)\dotfill &$2458345.8734\pm0.0013$\\
~~~~$a$\dotfill &Semi-major axis (AU)\dotfill &$0.0914\pm0.0017$\\
~~~~$i$\dotfill &Inclination (Degrees)\dotfill &$88.2^{+1.1}_{-1.0}$\\
~~~~$e$\dotfill &Eccentricity \dotfill &$0.3806^{+0.0093}_{-0.0090}$\\
~~~~$\omega_*$\dotfill &Argument of Periastron (Degrees)\dotfill &$57.1\pm1.7$\\
~~~~$T_{\rm eq}$\dotfill &Equilibrium temperature (K)\dotfill &$1198^{+18}_{-17}$\\
~~~~$M_{\rm P}$\dotfill &Mass (\mj)\dotfill &$5.42^{+0.22}_{-0.20}$\\
~~~~$K$\dotfill &RV semi-amplitude (m/s)\dotfill &$517.6\pm6.2$\\
~~~~$logK$\dotfill &Log of RV semi-amplitude \dotfill &$2.7140\pm0.0052$\\
~~~~$R_{\rm P}/R_*$\dotfill &Radius of planet in stellar radii \dotfill &$0.05588^{+0.00091}_{-0.00092}$\\
~~~~$a/R_*$\dotfill &Semi-major axis in stellar radii \dotfill &$11.09^{+0.28}_{-0.30}$\\
~~~~$\delta$\dotfill &Transit depth (fraction)\dotfill &$0.00312\pm0.00010$\\
~~~~$Depth$\dotfill &Flux decrement at mid transit \dotfill &$0.00312\pm0.00010$\\
~~~~$\tau$\dotfill &Ingress/egress transit duration (days)\dotfill &$0.01093^{+0.00085}_{-0.00050}$\\
~~~~$T_{\rm 14}$\dotfill &Total transit duration (days)\dotfill &$0.1964^{+0.0028}_{-0.0029}$\\
~~~~$T_{\rm FWHM}$\dotfill &FWHM transit duration (days)\dotfill &$0.1853\pm0.0029$\\
~~~~$b$\dotfill &Transit Impact parameter \dotfill &$0.22^{+0.12}_{-0.14}$\\
~~~~$b_{\rm S}$\dotfill &Eclipse impact parameter \dotfill &$0.43^{+0.23}_{-0.27}$\\
~~~~$\tau_S$\dotfill &Ingress/egress eclipse duration (days)\dotfill &$0.0229^{+0.0054}_{-0.0024}$\\
~~~~$T_{\rm S,14}$\dotfill &Total eclipse duration (days)\dotfill &$0.355^{+0.026}_{-0.043}$\\
~~~~$T_{\rm S,FWHM}$\dotfill &FWHM eclipse duration (days)\dotfill &$0.332^{+0.027}_{-0.048}$\\
~~~~$\delta_{\rm S,3.6\mu m}$\dotfill &Blackbody eclipse depth at 3.6$\mu$m (ppm)\dotfill &$115.2^{+7.8}_{-6.3}$\\
~~~~$\delta_{\rm S,4.5\mu m}$\dotfill &Blackbody eclipse depth at 4.5$\mu$m (ppm)\dotfill &$176.7^{+10.}_{-8.6}$\\
~~~~$\rho_{\rm P}$\dotfill &Density (cgs)\dotfill &$7.53^{+0.65}_{-0.72}$\\
~~~~$logg_{\rm P}$\dotfill &Surface gravity \dotfill &$4.162^{+0.026}_{-0.031}$\\
~~~~$\Theta$\dotfill &Safronov Number \dotfill &$0.908^{+0.030}_{-0.031}$\\
~~~~$\fave$\dotfill &Incident Flux (\fluxcgs)\dotfill &$0.407^{+0.026}_{-0.022}$\\
~~~~$T_{\rm P}$\dotfill &Time of Periastron (\bjdtdb)\dotfill &$2458326.549^{+0.025}_{-0.027}$\\
~~~~$T_{\rm S}$\dotfill &Time of eclipse (\bjdtdb)\dotfill &$2458323.485^{+0.052}_{-0.049}$\\
~~~~$T_{\rm A}$\dotfill &Time of Ascending Node (\bjdtdb)\dotfill &$2458325.863^{+0.025}_{-0.024}$\\
~~~~$T_{\rm D}$\dotfill &Time of Descending Node (\bjdtdb)\dotfill &$2458328.667^{+0.050}_{-0.049}$\\
~~~~$ecos{\omega_*}$\dotfill & \dotfill &$0.2065^{+0.0086}_{-0.0082}$\\
~~~~$esin{\omega_*}$\dotfill & \dotfill &$0.319\pm0.012$\\
~~~~$M_{\rm P}\sin i$\dotfill &Minimum mass (\mj)\dotfill &$5.42^{+0.22}_{-0.21}$\\
~~~~$M_{\rm P}/M_*$\dotfill &Mass ratio \dotfill &$0.004587^{+0.000100}_{-0.00011}$\\
~~~~$d/R_*$\dotfill &Separation at mid transit \dotfill &$7.17^{+0.24}_{-0.23}$\\
~~~~$P_{\rm T}$\dotfill &A priori non-grazing transit prob \dotfill &$0.1316^{+0.0044}_{-0.0043}$\\
~~~~$P_{\rm T,G}$\dotfill &A priori transit prob \dotfill &$0.1472^{+0.0050}_{-0.0048}$\\
~~~~$P_{\rm S}$\dotfill &A priori non-grazing eclipse prob \dotfill &$0.0677^{+0.0020}_{-0.0016}$\\
~~~~$P_{\rm S,G}$\dotfill &A priori eclipse prob \dotfill &$0.0757^{+0.0023}_{-0.0018}$\\
\smallskip\\\multicolumn{2}{l}{Wavelength Parameters:}&i'&z'&TESS\smallskip\\
~~~~$u_{1}$\dotfill &linear limb-darkening coeff \dotfill &$0.325\pm0.051$&$0.263\pm0.050$&$0.310\pm0.049$\\
~~~~$u_{2}$\dotfill &quadratic limb-darkening coeff \dotfill &$0.271^{+0.050}_{-0.051}$&$0.270^{+0.049}_{-0.050}$&$0.263^{+0.049}_{-0.050}$\\
~~~~$A_D$\dotfill &Dilution from neighboring stars \dotfill &$0.0017734^{+0.0000063}_{-0.0000062}$&$0.014331\pm0.000051$&$0.007830\pm0.000028$\\
\smallskip\\\multicolumn{2}{l}{Telescope Parameters:}&FEROS&TRES\smallskip\\
~~~~$\gamma_{\rm rel}$\dotfill &Relative RV Offset (m/s)\dotfill &$-6240.2^{+10.}_{-9.5}$&$-195.7\pm4.2$\\
~~~~$\sigma_J$\dotfill &RV Jitter (m/s)\dotfill &$30.0^{+12}_{-8.4}$&$6.4^{+7.7}_{-6.4}$\\
~~~~$\sigma_J^2$\dotfill &RV Jitter Variance \dotfill &$900^{+900}_{-430}$&$41^{+160}_{-95}$\\
\smallskip\\\multicolumn{2}{l}{Transit Parameters:}&LCO UT 2018-09-22 (i')&LCO UT 2018-10-11 (z')&TESS \smallskip\\
~~~~$\sigma^{2}$\dotfill &Added Variance \dotfill &$0.0000192^{+0.0000030}_{-0.0000025}$&$0.0000085^{+0.0000016}_{-0.0000013}$&$-0.0000000139^{+0.0000000083}_{-0.0000000078}$\\
~~~~$F_0$\dotfill &Baseline flux \dotfill &$1.00051\pm0.00046$&$0.99970^{+0.00038}_{-0.00037}$&$1.000007\pm0.000013$\\
\hline
\end{tabular}
 \begin{flushleft} 
  \footnotesize{ 
    \textbf{\textsc{\hspace{0.75in}NOTES:}}
$^\dagger$Minimum covariance with period.
All values in this table for the secondary occultation of \thisstar\ b are predicted values from our global analysis.               
               }
 \end{flushleft}
\label{tab:exofast}
\end{table*}

\section{Discussion}
\label{sec:discussion}
Our global analysis indicates that \thisstar\ has interesting characteristics that warrant further study. Specifically, \thisstar\ b is now one of only now four known planets that has a highly eccentric orbit ($>0.3$), a high planetary mass ($>$3 \mj), relatively short period ($<$20 days), and is bright enough ($V<$12) to be well suited for atmospheric characterization\footnote{\url{https://exoplanetarchive.ipac.caltech.edu/}; \citet{Akeson:2013}}. The host star has a mass of $M_{\star}$ = $1.128^{+0.065}_{-0.061}$ $M_{\odot}$, a radius of $R_{\star}$ = $1.777^{+0.047}_{-0.044}$, a surface gravity of $\log$ $g_{\star}$ = $3.993^{+0.027}_{-0.028}$, and an age of $7.4^{+1.6}_{-1.5}$ Gyr. Therefore, \thisstar\ appears to have just evolved off the main sequence and to be entering into the relatively short sub-giant phase (see Figure \ref{fig:discussion}). 

\subsection{Tidal Evolution and Irradiation History}
\label{sec:evo}
To gain a better understanding of the past and future evolution of \thisstar\ b's orbit, we use the latest version of POET\footnote{https://github.com/kpenev/poet}, where the results of our EXOFASTv2 global analysis (see \S\ref{sec:GlobalModel}) are used as boundary conditions. POET is a tool for calculating the evolution of a planetary orbit (circular-aligned) as a result of tidal dissipation (see \citealp{Penev:2014} for a detailed description of the original version of POET). Here we present an overview of the major changes that were used for the analysis of \thisstar\ b. The current version allows for inclined and eccentric orbits, where either object in the binary system can be a star or a planet. For the purposes of \thisstar, the difference between a star and a planet is that stars evolve (e.g. their radius changes) while planets do not. We assume that the star follows the MIST evolutionary tracks used in the EXOFASTv2 global model and that the rotation period of \thisstar\ is always longer than the orbital period of the planet. We note that this is not strictly true, since stars similar to \thisstar\ typically have a rotation period less than $\sim$9.5 days earlier in their lifetime while they are on the main-sequence. This assumption only affects the very early part of the analysis (near a zero-age main sequence), since the part of the evolution after the star has started spinning slower than the orbit is determined entirely by the present state of the system. We note that the estimated $\vsini$ from the TRES spectroscopy suggests a maximum rotation period of 17.6 days. 

\begin{figure}
\vspace{.0in}
\includegraphics[width=1\linewidth]{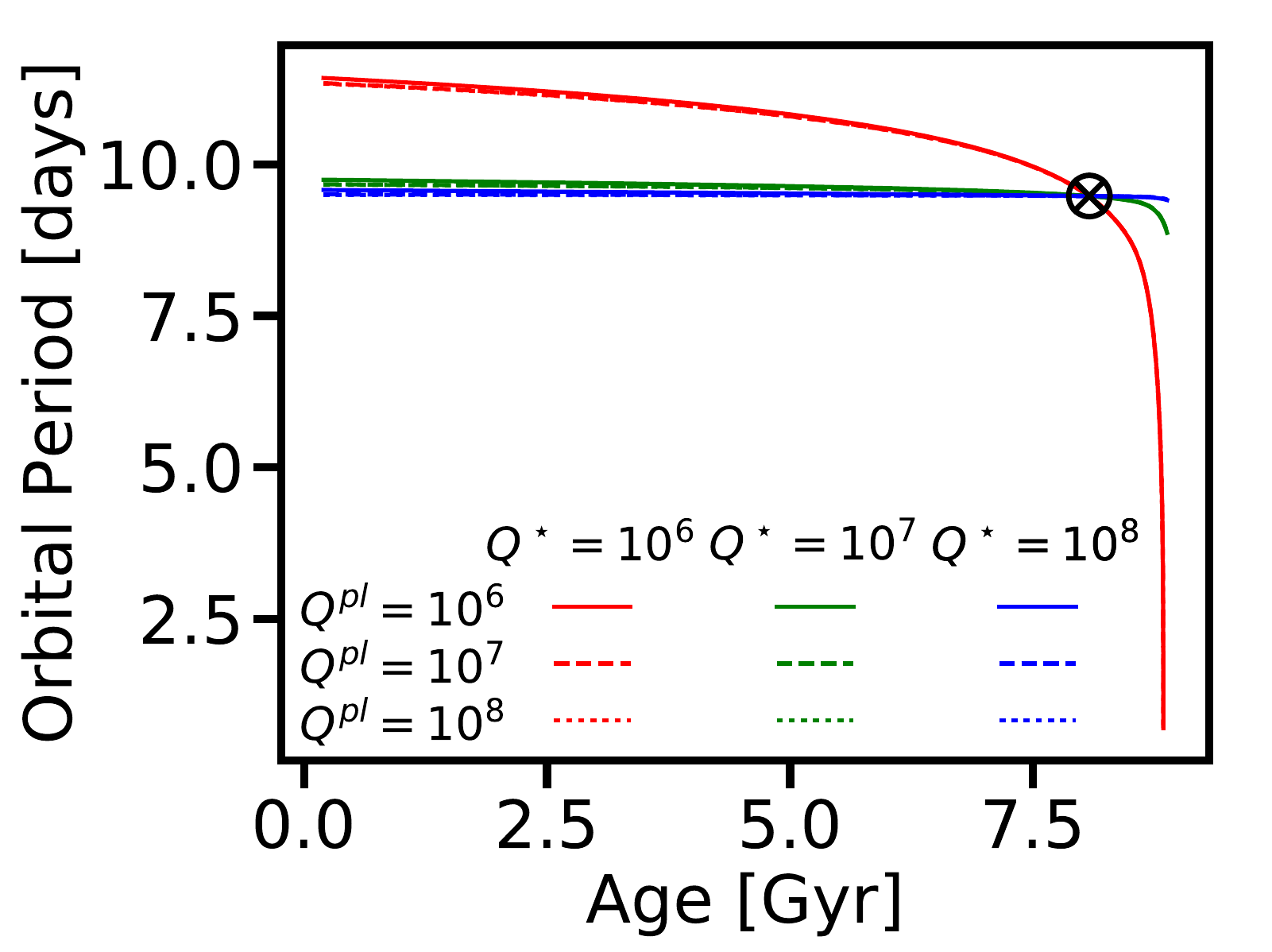}

\includegraphics[width=1\linewidth]{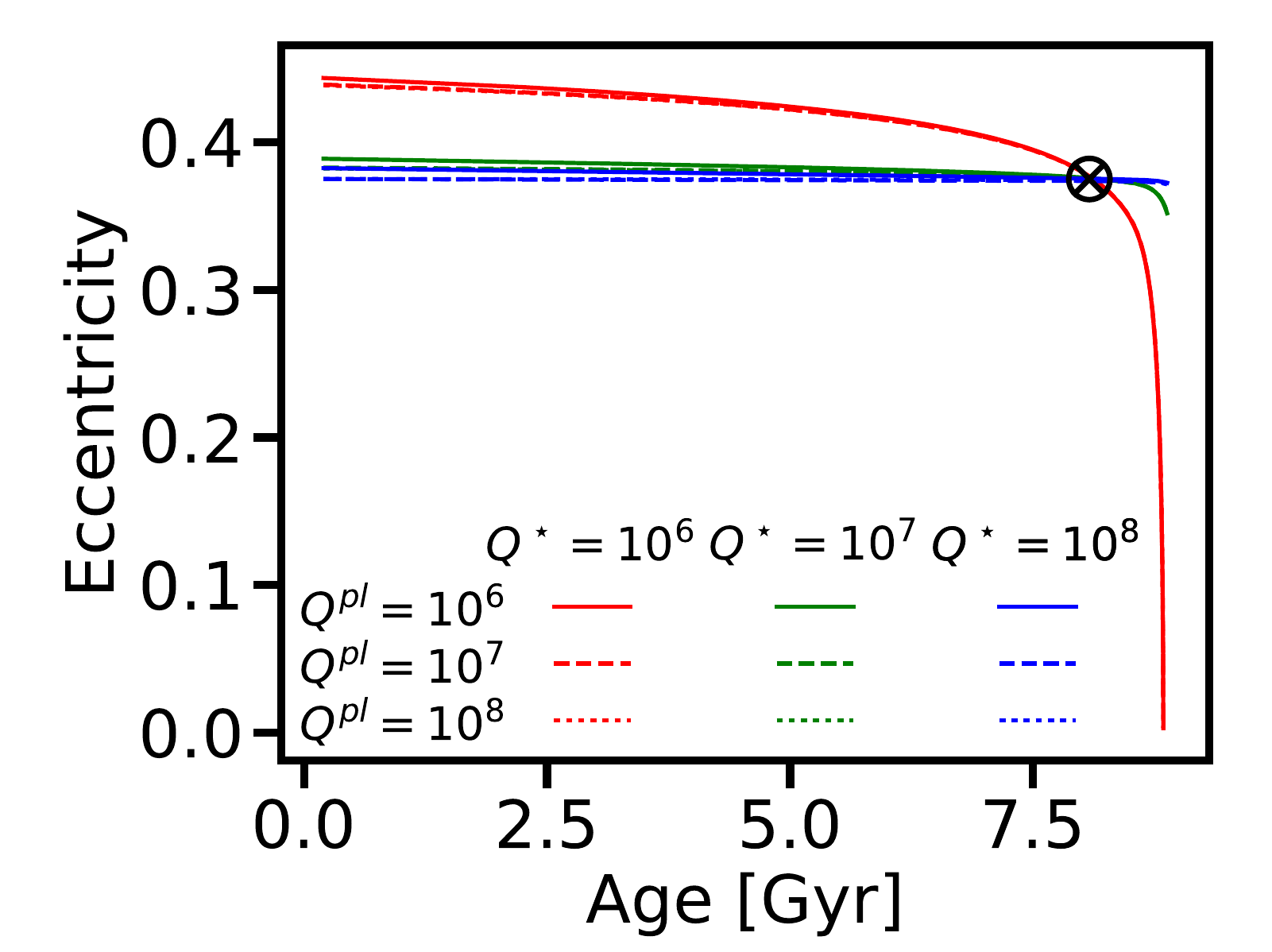}
\caption{Evolution of the  (\textit{Top}) orbital period and (\textit{bottom}) eccentricity for \thisstar\ b shown for a range of values for $Q^{\star}$. The color of the line indicates the dissipation in the star (red: Q$^{\star}=10^6$, green: Q$^{\star}=10^7$, blue: Q$^{\star}=10^8$) and the line style indicates the dissipation in the planet (solid: Q$_{P}=10^6$, dashed: Q$_{P}=10^7$ and dotted: Q$_{P}=10^8$). The tidal circularization in this system is dominated by tides raised on the star, rather than the planet (see \S\ref{sec:evo}).
}
\label{fig:evo} 
\end{figure}

For \thisstar, orbital evolutionary tracks were calculated for Q$^{\star}$= 10$^6$, 10$^7$ and 10$^8$, and for each of those,  Q$_P$ = 10$^6$, 10$^7$ and 10$^8$ (see Figure \ref{fig:evo}). The tidal quality factor (Q) defines the efficiency of tidal dissipation within the planet or star. Each track uses initial conditions that reproduce the present day orbital period and eccentricity of the system. Unfortunately, due to the high density of the planet and its relatively large semi-major axis, we are unable to produce meaningful constraints on Q$_P$ or Q$^{\star}$ (see Figure \ref{fig:evo}). In particular, even for Q$^{\star}$=10$^6$ and Q$_P$=10$^6$ the amount of circularization this system has undergone is relatively low. When using Q$^{\star}$ = 10$^5$, we are not able to find an initial eccentricity large enough to replicate the present eccentricity observed for \thisstar\ b. However, we are unable to try initial eccentricities larger than about e=0.6, because the Taylor series expansion of the tidal potential in eccentricity diverges past that point. 

This system contradicts normal conventional wisdom that tidal circularization is dominated by tides raised on the planet since the rate of circularization scales as (M'/M)$\times$R$^5$, where M and R are the mass and radius of the body experiencing the tides, and M' is the mass of the companion \citep{Adams:2006}. For a typical Jupiter mass planet around a Solar-type star, the contribution from tides raised on the planet is stronger than that from tides raised on the star by a factor of ten.
However, \thisstar\ b is more massive than Jupiter (M$_P$ = 5.4 \mj) and the host star is larger than the Sun (R$_{\star}$ = 1.78 R$_{\odot}$). Compared to the fiducial case, these contribute to an increase by a factor of nearly 90 in the rate of circularization due to tides raised on the star and a decrease by a factor of about 7 in the rate of circularization due to tides raised on the planet. Therefore, the present-day orbital evolution of \thisstar\ b is dominated by the tides raised on its host star by the planet. 

\subsection{Atmospheric Characterization Prospects}
The high eccentricity observed in the planet's orbit combined with the slight evolution of the host star make \thisstar\ b an interesting target for detailed characterization. While it is possible that hot Jupiters form in situ \citep{Batygin:2016}, most formation theories suggest that these planets form at larger distances from their host stars (core accretion or gravitational instability; \citealp{Pollack:1996, Lissauer:2007, Boss:2000, Gammie:2001, Boley:2009}) and migrate inward via two main interactions, either with drag due to the original protoplanetary disk during formation, or by gravitational interaction with another body in the system \citep{Rasio:1996, Papaloizou:2007, Fabrycky:2007}. It was originally believed that the large number of hot Jupiters shown to have misaligned orbits relative to the host star's spin axis indicated that these systems must migrate through gravitational scattering \citep{Winn:2010b}. However, the origin of these misalignments could have occurred from misalignments in the protoplanetary disk \citep{Batygin:2012, Crida:2014}. Therefore, it is unclear what migration mechanism is responsible for close-in Jovian planets. 

The migration mechanism may be revealed by studying the chemical abundances in the planet's atmosphere. Specifically, it is more difficult to explain low carbon and oxygen abundances relative to the planet's host star via disk migration than via disk-free migration \citep{Madhusudhan:2014}. 
The high eccentricity of \thisstar\ b is suggestive of disk-free migration, although our investigation of its orbital evolution suggests it never possessed the extremely high eccentricity that would be required to migrate from a formation location beyond the ice line. Moreover, as discussed previously, it is plausible that many hot Jupiters migrated first in the disk and then consequently through planet-planet interactions; even eccentric planets may have disk migration in their history. Nonetheless, there are currently only about a dozen planets larger than Neptune for which the eccentricity is greater than 0.2 with at least 99\% confidence, a relatively short period (<20 days), and are bright enough ($V<12$) to be well-suited for atmospheric characterization\footnote{\url{https://exoplanetarchive.ipac.caltech.edu/} ; \citet{Akeson:2013}}. Interestingly, less than half of these systems (including \thisstar) have a massive planetary companion ($>$3 \mj).  Therefore, \thisstar\ b---with the other few known planets in this sub-sample, such as HAT-P-2b \citep{Bakos:2007} and WASP-162 \citep{Hellier:2019}---provides a great opportunity to carry out this test. If a depletion of oxygen and carbon are detected, it could provide evidence that it migrated via a disk-free method, or otherwise place constraints on its disk migration history. Future observations could try to characterize the composition of \thisstar\ b's atmosphere using current facilities like the {\it Hubble Space Telescope} and future facilities like the {\it James Webb Space Telescope}. Additionally, understanding the full architecture of the \thisstar\ system, by looking for long period giant planet companions through radial velocity monitoring, may provide additional insight into its evolutionary history.

\section{Conclusion}
\label{sec:conclusion}
We present the discovery of \thisstar\ b, a massive Jupiter in a highly eccentric $\sim$9.5 day orbit around a slightly evolved G-star. The planet has a very high density (M$_P$ = $5.42^{+0.22}_{-0.20}$ \mj, R$_P$ = $0.965^{+0.032}_{-0.029}$ \rj, $\rho_P$ = $7.53^{+0.65}_{-0.72}$ g cm$^{-3}$) while its host star appears to be a sub-giant (M$_{\star}$ = $1.128^{+0.065}_{-0.061}$ \msun, R$_{\star}$ = $1.777^{+0.047}_{-0.044}$ \rsun, \logg = $3.993^{+0.027}_{-0.028}$). Interestingly, \thisstar\ b is in a rare class of highly eccentric (>0.3), short-period ($<$20 days) massive ($>$3 \mj) planets. The large mass and semi-major axis of \thisstar\ b corresponds to a circularization timescale much larger than the age of the universe. The large eccentricity of the planet's orbit suggests that at least some of its migration history included dynamical interactions with other components in the system. From studying the orbital evolutionary history of \thisstar\, we are unable to place any useful constraints on Q$_{P}$ or Q$^{\star}$ since the tidal evolution is expected to be slow in this system for all reasonable values of Q$^{\star}$ and Q$_P$. Future observations could provide more evidence for the migration mechanism by studying the atmospheric composition of \thisstar\ b or by studying the entire known ensemble of hot Jupiters in the literature. 


\software{EXOFASTv2 \citep{Eastman:2013, Eastman:2017}, AstroImageJ \citep{Collins:2017}}
\facilities{TESS, FLWO 1.5m (Tillinghast Reflector Echelle Spectrograph), 4.1-m Southern Astrophysical Research (SOAR), LCO 0.4m, LCO 1.0m, 2.2m telescope La Silla (Fiber-fed Extended Range Optical Spectrograph)}

\acknowledgements

We thank Laura Kreidburg and Laura Mayorga for their valuable conversations. J.E.R. was supported by the Harvard Future Faculty Leaders Postdoctoral fellowship. AV's contribution to this work was performed under contract with the California Institute of Technology (Caltech)/Jet Propulsion Laboratory (JPL) funded by NASA through the Sagan Fellowship Program executed by the NASA Exoplanet Science Institute. CZ is supported by a Dunlap Fellowship at the Dunlap Institute for Astronomy \& Astrophysics, funded through an endowment established by the Dunlap family and the University of Toronto. CXH, JAB, and MNG acknowledge support from MIT’s Kavli Institute as Torres postdoctoral fellows. DD acknowledges support for this work provided by NASA through Hubble Fellowship grant HST-HF2-51372.001-A awarded by the Space Telescope Science Institute, which is operated by the Association of Universities for Research in Astronomy, Inc., for NASA, under contract NAS5-26555. TD acknowledges support from MIT’s Kavli Institute as a Kavli postdoctoral fellow. R.B. acknowledges support from FONDECYT Post-doctoral Fellowship Project 3180246, and from the Millennium Institute of Astrophysics (MAS). A.J.\ acknowledges support from FONDECYT project 1171208, CONICYT project BASAL AFB-170002, and by the Ministry for the Economy, Development, and Tourism's Programa Iniciativa Cient\'{i}fica Milenio through grant IC\,120009, awarded to the Millennium Institute of Astrophysics (MAS).

This research has made use of SAO/NASA's Astrophysics Data System Bibliographic Services. This research has made use of the SIMBAD database, operated at CDS, Strasbourg, France. This work has made use of data from the European Space Agency (ESA) mission {\it Gaia} (\url{https://www.cosmos.esa.int/gaia}), processed by the {\it Gaia} Data Processing and Analysis Consortium (DPAC, \url{https://www.cosmos.esa.int/web/gaia/dpac/consortium}). Funding for the DPAC has been provided by national institutions, in particular the institutions participating in the {\it Gaia} Multilateral Agreement. This work makes use of observations from the LCO network

Funding for the {\it TESS} mission is provided by NASA's Science Mission directorate. We acknowledge the use of public {\it TESS} Alert data from pipelines at the {\it TESS} Science Office and at the {\it TESS} Science Processing Operations Center. This research has made use of the NASA Exoplanet Archive and the Exoplanet Follow-up Observation Program website, which are operated by the California Institute of Technology, under contract with the National Aeronautics and Space Administration under the Exoplanet Exploration Program. This paper includes data collected by the {\it TESS} mission, which are publicly available from the Mikulski Archive for Space Telescopes (MAST). This paper includes observations obtained under Gemini program GN-2018B-LP-101. Resources supporting this work were provided by the NASA High-End Computing (HEC) Program through the NASA Advanced Supercomputing (NAS) Division at Ames Research Center for the production of the SPOC data products.


\bibliographystyle{apj}

\bibliography{TOI172}



\end{document}